\documentclass[pra,showpacs,preprintnumbers,amsmath,amssymb]{revtex4-1}
\usepackage{graphicx}
\usepackage{graphics}
\usepackage{amsmath,amsfonts,amssymb}
\usepackage{epsfig}
\usepackage{color}

\begin{document}


\title{Explicit schemes for time propagating many-body wavefunctions}

\author{Ana Laura Frapiccini$^1$\footnote{On leave of absence from CONICET and Departamento de F\'{i}sica, Universidad Nacional 
                                                                              del Sur, 8000 Bahia Blanca, Argentina.},  
              Aliou Hamido$^1$, 
              Sebastian Schr\"oter$^2$,
              Dean Pyke$^3$, Francisca Mota-Furtado$^3$, Patrick F. O'Mahony$^3$,
              Javier  Madro\~{n}ero$^4$, Johannes  Eiglsperger$^5$ and 
              Bernard  Piraux$^1$}

\affiliation{$^1$Institute of Condensed Matter and Nanosciences (IMCN),  Universit\'e Catholique de  Louvain,\\
                            B\^{a}timent de Hemptinne,  2, chemin du cyclotron, B1348 Louvain-la-Neuve, Belgium.\\
                   $^2$Physik Department, Technische Universit\"at M\"{u}nchen, D-85747 Garching,  Germany.\\
                   $^3$Department of Mathematics, Royal Holloway, University of London, Egham,\\
                             Surrey TW20 0EX, United Kingdom.\\
                   $^4$Departamento de F\'{i}sica, Universidad del Valle, A.A. 25360, Cali, Columbia.\\
                   $^5$numares GmbH, Josef-Engert-Stra\ss e 9, D-93053 Regensburg, Germany.}
\date{\today}

\begin{abstract}
Accurate theoretical data on many time-dependent processes in atomic and molecular physics and in chemistry require the direct numerical {\it ab initio} solution of the time-dependent Schr\"odinger equation, thereby motivating the development of very efficient time propagators. These usually involve the solution of very large systems of first order differential equations that are characterized by a high degree of stiffness. In this contribution, we analyze and compare the performance of the explicit one-step algorithms of  Fatunla and Arnoldi. Both algorithms have exactly the same stability function, therefore sharing the same stability properties that turn out to be optimum. Their respective accuracy however differs significantly and depends on the physical situation involved. In order to test this accuracy, we use a predictor-corrector scheme in which the predictor is either Fatunla's or Arnoldi's algorithm and the corrector, a fully implicit four-stage Radau IIA method of order 7. In this contribution, we consider two physical processes. The first one is the ionization of an atomic system by a short and intense electromagnetic pulse; the atomic systems include a one-dimensional Gaussian model potential as well as atomic hydrogen and helium, both in full dimensionality. The second process is  the decoherence of two-electron quantum states  when a  time independent perturbation is applied to a planar two-electron quantum dot where both electrons are confined in an anharmonic potential. Even though the Hamiltonian of this system is time independent the corresponding differential equation shows a striking stiffness which makes the time integration extremely difficult. In the case of the one-dimensional Gaussian potential we discuss in detail the possibility of monitoring the time step for both explicit algorithms. In the other physical situations that are much more demanding in term of computations, we show that the accuracy of both algorithms depends strongly on the degree of stiffness of the problem. 
\end{abstract}

\maketitle


\section{Introduction}
\label{intro}

The numerical integration of the time-dependent Schr\"odinger equation (TDSE) has become the main theoretical approach for the quantitative study of a vast amount of phenomena, including strong field processes in atoms and molecules, quantum collisions and chemical reactions. In strong field physics, current  light sources can create ultrashort pulses of very high intensity making the numerical solution of the TDSE unavoidable if accurate results are required. In the low frequency regime where the photon energy is much lower than the ionization potential, the advent of high-intensity lasers has allowed detailed investigations of phenomena such as above-threshold ionization \cite{Muller}, high order harmonic generation \cite{L'Huillier}, multiphoton multiple ionization \cite{Hansch}, attosecond pulse generation \cite{Antoine}, molecular self-spectroscopy \cite{Corkum}, etc. In the high frequency regime where the photon energy is of the order or larger than the ionization potential, very intense coherent X-ray sources  are under development. They are based on the collective electronic response of a plasma to ultra intense laser fields \cite{Borot} as well as the next generation free electron lasers (FEL) such as the European XFEL project. The latter  is expected to boost the average photon flux by about two orders of magnitude in comparison with  already existing FELs. The interaction of atoms or molecules with intense X-ray pulses with a duration in the femtosecond or subfemtosecond regime is expected to lead to highly non-linear processes which can no longer be described within perturbation theory as is currently the case.\\

In quantum collision theory, the interaction Hamiltonians do not usually depend explicitly on time. Time independent approaches such as the R-matrix \cite{Burke} or the S-matrix  methods \cite{Joachain} suffice. However, in many cases, the lack of knowledge of the asymptotic boundary conditions or the explicit introduction of a time in the interaction Hamiltonian through the classical description of a heavy projectile makes the numerical solution of the corresponding TDSE more convenient \cite{Sidky}. Methods such as the time-dependent close coupling \cite{Pindzola} are particularly efficient. Nevertheless, when the quantum systems involved become more complex as is often the case for chemical reactions or in condensed matter physics, the numerical solution of the TDSE is no longer possible. Different approaches are necessary as for instance the time-dependent density functional theory (TDDFT)\cite{Runge, Marques}. In that case, the Hamiltonian is replaced by the self-consistent Kohn-Sham Hamiltonian. In fact, TDDFT can be viewed as a reformulation of time-dependent quantum mechanics where the basic unknown is no longer the many-body wavefunction, but the time-dependent electron density \cite{Castro}. This density can be obtained from the solution of a set of one-body equations, the so-called Kohn-Sham equations that have the same form as the usual TDSE.\\

The necessity of integrating numerically the TDSE motivates the development of efficient and accurate time pro\-pagators. Once the TDSE has been discretized in the spatial or/and energy domain by means of either a finite difference grid method or an approach based on spectral or finite element methods, the time integration of the TDSE reduces to the solution of a system of  first order differential equations which may be written as follows:
\begin{equation}
\frac{\mathrm{d}}{\mathrm{d}t}\mathbf{Y}=-\mathrm{i}\mathbf{H}(t)\mathbf{Y},\label{eq1}
\end{equation}
where $\mathbf{H}(t)$ is a matrix that depends explicitly on time. The main difficulty we have to face in solving such a system of ordinary differential equations is the fact that the spectrum of matrix $\mathbf{H}$ is not bound. In general, the matrix $-\mathrm{i}\mathbf{H}$  has very high, purely imaginary, eigenvalues. These very high eigenvalues  give rise to extremely fast oscillations of the true solution and usually determine the time step of the numerical time propagator. Another way to describe this problem is to say that the system  behaves as a stiff system. Although there is no rigorous mathematical definition of the stiffness, a system is said to be stiff in a given interval of integration if the numerical method is forced to use a step length which is excessively small in relation to the smoothness of the exact solution \cite{Lambert}. In addition, increasing the size of the system generates eigenvalues each time higher thereby increasing its stiffness.\\

The problem associated with stiff systems is twofold: stability and accuracy. To each numerical method is associated a function named the stability function that determines the stability properties of the method and the range of time steps for which the numerical solution is stable and remains bounded. In the case of stiff systems, it can be shown, for instance, that none of the explicit methods of Runge Kutta  (R-K) type is stable. In that case, it is necessary to use an implicit R-K scheme. Note that, by contrast to an explicit method which only requires matrix-vector products, all implicit schemes require solving systems of algebraic equations at each time step. For systems of considerable size, the computer time becomes, in these conditions, rapidly excessive. Fortunately, explicit schemes exist that are not of R-K type but having the stability properties required for dealing with such stiffness problems. The accuracy problem is more delicate. If an appropriate integration method is used, the stability problem may be avoided but, for a reasonable step length, the solution components corresponding to the largest eigenvalues are approximated very inaccurately \cite{Lapidus}. However, there is no mathematical tool which allows one to predict whether the numerical solution of a stiff system will be accurate or not. Very often, the highest eigenvalues that correspond to very high energies do not play any physical role but, this does not imply that the error made in calculating the corresponding high energy components of the full numerical solution will not affect the final result. In fact, it is important to proceed on a case by case basis.\\

In this contribution, we analyze in detail two explicit one-step integration schemes that have the required stability properties for dealing with stiff systems. The first method is due to Fatunla \cite{Fatunla1, Fatunla2} and the second one is a Krylov subspace method usually called the Arnoldi algorithm. The Arnoldi algorithm has already been used in many different contexts: strong field physics \cite{Smyth}, condensed matter physics \cite{Castro}, etc...  However, as far as we know, no systematic study of its stability and accuracy properties exists so far. In order to test the accuracy of both methods, we use a predictor-corrector scheme in which the predictor is either Fatunla's method or Arnoldi's algorithm while the corrector is a four-stage diagonally implicit Radau IIA method of order seven. Here, we consider the interaction of a quantum system with a strong and ultrashort electromagnetic pulse and test the three methods in the case of three different quantum systems: a model potential, atomic hydrogen and helium. We also examine the performances of these explicit schemes in a completely different context namely the calculation of a fidelity function that measures the decoherence of two-electron quantum states  when a  time independent perturbation is applied to a planar two-electron quantum dot where both electrons are confined in an anharmonic potential. In fact this is a difficult problem, which exhibits a strong degree of stiffness although its Hamiltonian is time independent. Finally, let us mention that a comparison of different time propagation algorithms for the time dependent Schr\"odinger equation may be found in \cite{Leforestier}.\\

This article is organized as follows. Section II is devoted to the general formulation of the TDSE. After some preliminary remarks, we give and discuss the general spectral representation of the TDSE and finally define the stability function of a given algorithm. In section III, we introduce the various algorithms (Fatunla's method, Arnoldi's algorithm and the predictor-corrector scheme) in the context of our model potential. For both explicit schemes, we give  their stability function and analyze in detail their accuracy in the case of the model potential. Section IV is devoted to the results obtained with Fatunla's and  Arnoldi's methods for the model potential, the interaction of atomic hydrogen with both a high and a very low frequency strong laser field and single ionization of helium. Finally, we consider the problem of the planar two-electron quantum dot. Unless otherwise stated, atomic units are used throughout this paper.\\


\section{General formulation of the TDSE}
\label{tdse}

\subsection{Preliminary remarks}
\label{preliminary}

Our aim is to study the interaction of a quantum system with an external time-dependent field. Solving numerically the corresponding TDSE proceeds in two steps: the discretization in the spatial or/and energy domain of the equation and the time propagation of the solution. There are typically three ways of discretizing the TDSE: the finite difference grid (FDG) methods and the approaches based on spectral or finite element methods.  The simplest approaches are the FDG methods. These methods based on a spatial discretization are essentially local. They are very often used because the subsequent time propagation involves solving very sparse systems of algebraic equations. However, it is often tricky to extract information on how these methods account for the electronic structure of the quantum system under consideration and some observables are sometimes difficult to calculate. Furthermore, these methods yield finite-order rates of convergence in terms of the number of spatial grid points. In other words, the errors go to zero as the inverse of this number at a power given by the order of the method.\\

The spectral methods based on an energy discretization are non-local. They consist in writing the solution as a truncated expansion in terms of $\mathcal{L}^2$ integrable functions. These functions  form a complete basis set. Different choices of basis sets are possible, which usually depend on the physics of the problem. The most commonly used functions are the Hermite functions, the Coulomb Sturmian functions \cite{Rotenberg} and orthogonal polynomials. There are essentially two types of spectral methods: Galerkin and collocation \cite{Gottlieb}. In addition to the energy discretization as is the case in the Galerkin method, the collocation method involves also a spatial discretization. However, by contrast to the FDG methods, the grid mesh points are not arbitrarily chosen. They are the abscissae of the Gaussian quadrature associated with the basis functions. The spectral approaches are very appropriate for a very accurate description of the bound and resonance states of the quantum system under consideration. This is particularly true for resonance states very close to the ionization thresholds \cite{Eiglsperger}. The convergence of the spectral methods in terms of the number of basis functions depends on the analytical properties of the solution. If the successive spatial derivatives of this solution do not exhibit singularities, the convergence is exponential. This means that the errors go to zero faster than any finite power of the inverse of the number of basis functions. On the other hand, if the solution exhibits singularities in its successive derivatives and if the basis wavefunctions do not account for these singularities,  the convergence is much slower. Typical examples of such singularities are the Kato cusps present in many-particle system wavefunctions \cite{Kato}. Another drawback of the spectral methods is the fact that the matrix associated with the Hamiltonian is, in most cases, not sparse. \\\\

The finite element methods which are based on a subdivision of the whole spatial domain of integration into simple subdomains are in fact closely related to the spectral methods. They differ, however, by the fact that the basis functions have  bounded support, being  therefore piecewise regular. In addition, these methods yield also finite-order rates of convergence like in the case of the FDG methods. Piecewise Lagrange polynomials or B-splines are very often used as basis functions. In general, these methods are particularly efficient in describing the electronic continuum states of the system under consideration. In addition, singularities or large gradients in the solution  can be treated by considering non-regular subdomains. These methods are very often used, especially those based on B-splines \cite{Bachau} because the subsequent time propagation involves relatively sparse systems of equations to solve like in the case of the FDG methods. In the present contribution, we use spectral or/and B-spline based methods in all the cases treated. \\

\subsection{The spectral representation of the TDSE}
\label{spectral}

The TDSE for a quantum system interacting with an external field can be written as
\begin{equation}
\mathrm{i}\,\frac{{\partial\Psi \left( {\mathbf{r},t} \right)}}{{\partial t}} = H\left( {\mathbf{r},t} \right)\,\Psi \left( {\mathbf{r},t} \right),\label{eq2}
\end{equation}
where $\Psi \left( {\mathbf{r},t} \right)$ is the wavefunction of the system, $\mathbf{r}$ represents any set of $n$ spatial coordinates and $t$ is the time.  The total Hamiltonian $H\left( {\mathbf{r},t} \right)$, which depends explicitly on time, is given by
\begin{equation}
H\left( {\mathbf{r},t} \right)=H_0\left( {\mathbf{r}} \right)+ V\left( {\mathbf{r},t} \right),\label{eq3}
\end{equation}
with $H_0\left( {\mathbf{r}} \right)$ the unperturbed Hamiltonian and $V\left( {\mathbf{r},t} \right)$ the time-dependent interaction potential (velocity form in all the cases treated here).  Using a complete basis set $\left\{ {{f_i}\left(\mathbf{ r} \right)} \right\}$ of square integrable functions, we write the wavefunction $\Psi(\mathbf{r},t)$, the solution of equation (\ref{eq2}), as the following truncated expansion,
\begin{equation}
\Psi(\mathbf{r},t)=\sum_{i=1}^{N}\psi_{i}(t)f_{i}(\mathbf{r}), \label{eq4}
\end{equation}
where the expansion coefficients $\psi_{i}(t)$ are time-dependent. $N$ represents the number of terms in the expansion and is taken sufficiently large to represent the wavefunction to the desired accuracy. As a result, the TDSE is transformed into a matrix equation for the vector $\boldsymbol{\Psi}(t)=\{\psi_{i}(t)\}_{N}$, given by
\begin{equation}
\mathrm{i}\; \mathbf{B} \frac{\mathrm{d}}{\mathrm{d} t}\boldsymbol{\Psi}(t)=\mathbf{H}(t) \boldsymbol{\Psi}(t). \label{eq5}
\end{equation}
For a non-orthonormal basis, the overlap matrix $\mathbf{B}$ and the Hamiltonian $\mathbf{H}(t)$ have elements defined by
\begin{eqnarray}
\left[\mathbf{B}\right]_{ij}&=&\langle f_{i} | f_{j} \rangle, \label{eq6}\\
\left[\mathbf{H}\right]_{ij}&=&\langle f_{i} | H(\mathbf{r},t) | f_{j}\rangle. \label{eq7}
\end{eqnarray}
The time evolution of the wavepacket is then given by the solution of the following $N$-dimensional system of first order differential equations:
\begin{equation}
\frac{\mathrm{d}}{\mathrm{d}t}\boldsymbol{\Psi}(t)=-\mathrm{i}\; \mathbf{B}^{-1}\mathbf{H}(t)\boldsymbol{\Psi}(t). \label{eq8}
\end{equation}
There is actually no need to evaluate explicitly the inverse of the overlap matrix $\mathbf{B}$.  This matrix is always symmetric and positive definite, which allows a numerically stable and fast Cholesky decomposition.  In that case the action of $\mathbf{B}^{-1}$ on a vector can be calculated straightforwardly by solving a very sparse system of algebraic equations. The vector $\boldsymbol{\Psi}(t)$ is said to be $\mathbf{B}$-orthogonal and its norm is given by
\begin{equation}
\boldsymbol{\Psi}^{\dagger} \cdot \mathbf{B} \cdot \boldsymbol{\Psi}=1. \label{eq9}
\end{equation}
Note that in the case of the FDG methods, a system of equations similar to the system (5) has to be solved but the matrix $\mathbf{H}$ is no longer associated with the Hamiltonian. \\

\subsection{The boundary and asymptotic conditions}
\label{asymptotic}

In solving numerically the TDSE, the discretization method has to account correctly for the non-trivial problems of the boundary and asymptotic conditions. By way of illustration, let us consider the case of the ionization of atomic hydrogen by an intense low frequency laser field. The amplitude of the electron quiver motion determines the minimum  spatial grid size or the minimum number of basis functions to be included. For high intensities and very low frequencies, this amplitude may become of the order of thousands of atomic units thereby requiring excessively long computational times. In addition, during the interaction process,  ionization takes place and fast emitted electrons will rapidly reach  the boundaries of the computational domain. It is therefore important to choose appropriate boundary conditions to avoid spurious reflections of the wavefunction at these boundaries. Such reflections can be avoided by further increasing the size of the computational domain, but this becomes rapidly untractable.  Instead, reflection problems can be overcome by introducing complex absorbing potentials \cite{DiMenza,Muga}. Those potentials however are usually not completely reflection free. A better approach is exterior complex scaling (ECS) in which the outgoing electron coordinate becomes complex beyond a certain distance from the nucleus which is larger than the amplitude of the quiver motion \cite{He,Scrinzi1}. \\

For single electron systems, the extraction of the information on the differential probability densities does not cause any problem since the asymptotic behavior of the field-free continuum states is known. This contrasts with the multi-electron systems where the asymptotic behavior of the multiple continuum wavefunctions is unknown. In that case, one can either develop approximate expressions for these continuum wavefunctions or use more sophisticated time-dependent methods that circumvent the problem. When the outgoing electrons are sufficiently far from each other so that their interaction becomes negligible, multiple continuum wavefunctions are usually approximated by a product of Coulomb functions \cite{Colgan,Laulan,Foumouo,Feist}. The validity of this approximation which gives  reliable results is discussed in \cite{Malegat1}. More sophisticated methods that avoid any projection of the final wavepacket on approximated multiple continuum wavefunctions have been developed. Palacios {\it et al.} \cite{Palacios} have derived a time-dependent method where the extraction of the information from the wavepacket is based on ECS. Malegat  {\it et al.} extract the information from the total wavepacket after propagating semiclassically its Fourier components in space over very large distances \cite{Malegat2,Malegat1}. Scrinzi has extended the time-dependent surface flux method to single and double ionization of two-electron systems \cite{Scrinzi2}. Hutchinson {\it et al.} \cite{Hutchinson} are developing a time-dependent R-matrix approach that can describe the interaction of any (light) atomic systems with short electromagnetic pulses. More recently, Hamido {\it et al.} have developed the so-called time scaled coordinate  (TSC) method \cite{Hamido}. This latter method which is used in some of the cases treated in this contribution, consists in performing a time-dependent scaling of the radial coordinates of the electrons together with a phase transformation of the wavefunction.  As a result, an harmonic potential appears in the scaled Hamiltonian, which confines the wavefunction in configuration space.  It can be shown that  a relatively long time after the interaction, the wavefunction becomes stationary and its modulus gives directly the momentum distribution of the particles resulting from the fragmentation of the system.  Consequently this method clearly circumvents  the above mentioned difficulties.  It however introduces different length scales that need to be treated with multiresolution techniques and that influence the stability of the numerical time propagation scheme.\\

\subsection{The stability function}
\label{stability}

In order to analyze the stability of a one-step numerical time propagation scheme, it is convenient to consider the following standard test problem (Dahlquist's equation):
\begin{equation}
\frac{\mathrm{d}y}{\mathrm{d} t}=\lambda y, \label{eq10}
\end{equation}
where $\lambda$ is a constant. If we assume that $y(0)=\eta$, the solution of this equation is $y(t)=\eta\exp(\lambda t)$. Usually, a system of equations is said to be stiff when its Jacobian matrix has some eigenvalues with a very large negative real part. In the case of Eq.(\ref{eq10}), assuming that the real part of $\lambda$ is very large and negative leads to a solution that tends extremely rapidly to zero. We have therefore to look for the conditions that have to be imposed on the numerical time propagation scheme in order that the numerical solution $y_n=y(n\delta t)\rightarrow 0$ as $n\rightarrow\infty$ where $\delta t$ is the time step. By applying the one-step numerical time propagation scheme to Eq.(\ref{eq10}), we obtain 
\begin{equation}
y_{n+1}=R(\lambda\delta t)y_n\label{eq11},
\end{equation}
where $R(z)$ is the so-called  stability function. In order that $y_n$ tends to zero as $n\rightarrow\infty$, we must impose  $R(\lambda\delta t) < 1$ thereby implying some constraints on the time step $\delta t$. The set $S=\{z=\lambda\delta t\in\mathbb{C};\;|R(z)|\leq1\}$ is called the stability domain of the numerical scheme. This latter one is said to be A-stable if its stability domain is included in $\mathbb{C}^-=\{z;\;Re\;z\leq 0\}$. It is L-stable if, apart from being A-stable, the stability function has the property $\lim_{Re(\lambda\delta t)\rightarrow -\infty}|R(\lambda\delta t)|=0$. L-stable methods are the most stable ones \cite{Lambert}. \\

In the present case, the Jacobian of the system of equations we are interested in has large purely imaginary eigenvalues. Although such systems behave like a stiff system, the analysis of the stability of the numerical scheme is more delicate. Suppose for instance that the numerical scheme we use is L-stable and that its stability domain covers the half-plane $\mathbb{C}^-$ as well as large parts of the right half-plane $\mathbb{C}^+$. In these conditions, uninteresting high oscillations of the true solution may be damped by the numerical scheme. However, the norm of the solution will not be necessarily preserved since $|R(\lambda\delta t)|\leq 1$. We must impose, as an additional constraint, that $|R(\lambda\delta t)|=1$. This means that if $\lambda$ is purely imaginary, $R(\lambda\delta t)=\exp(\lambda\delta t)$. Following Fatunla \cite{Fatunla2}, a numerical time propagation scheme is said to be exponentially fitted at a complex value $\lambda=\lambda _0$ if the stability function $R(\lambda\delta t)$ satisfies the relation
\begin{equation}
R(\lambda_0\delta t)=\exp(\lambda_0\delta t)\label{eq12}.
\end{equation}

\section{Time propagation algorithms} 
\label{algorithm}

In this section we describe and compare the performance of two explicit one-step time propagation schemes, namely Fatunla's method and Arnoldi's algorithm in terms of stability and accuracy. To test the accuracy of both schemes we use an implicit predictor-corrector (P-C) method. The predictor is either Fatunla's method or Arnoldi's one and the corrector is a four-stage Radau II-A implicit method which is of Runge-Kutta type.  Monitoring the time step during the time propagation using both explicit schemes is a key point which will be addressed first within a simple one-dimensional model of one electron in a Gaussian potential of the form $V(x)=-V_{0}e^{-\beta x^2}$, where $ V_{0}$ and $\beta$ are constants and exposed to a cosine square pulse.\\

In the following sections, these two algorithms will be also tested in two more demanding physical situations.  The first situation is the interaction of a quantum system with a strong electromagnetic pulse.  The quantum systems we shall be studying in that case are  atomic hydrogen and helium, both treated in full dimensions.  The second physical situation is the time evolution of a two-electron wavepacket in a two-dimensional quantum dot.

\subsection{Fatunla's method}
\label{fatunla}

The idea behind Fatunla's method is to take into account the intrinsic frequencies of the atom-field system by introducing interpolating oscillatory functions that approximate the solution of the TDSE.  This allows one to deal with problems displaying eigenvalues that differ by many orders of magnitude.  That explains why Fatunla's method has the capability to solve stiff equations, while requiring only matrix vector products.  More precisely, we write the first order differential equation (\ref{eq8}) as
\begin{equation}
\frac{\mathrm{d}}{\mathrm{d}t}\boldsymbol{\Psi}(t)=-\mathrm{i}\; \mathbf{B}^{-1}\mathbf{H}(t)\boldsymbol{\Psi}(t)=\mathbf{f}(t,\boldsymbol{\Psi})\label{eq13}.
\end{equation}
The solution $\boldsymbol{\Psi}(t)$ over a subinterval $[t_{n},t_{n}+\delta t=t_{n+1}]$ is approximated by the interpolating oscillatory function
\begin{equation}
\widetilde{\mathbf{F}}(t)=(\mathbf{I}-e^{\boldsymbol{\Omega}_{1}t})\mathbf{a}-(\mathbf{I}-e^{-\boldsymbol{\Omega}_{2}t})\mathbf{b}+\mathbf{c}, \label{eq14}
\end{equation}
with $\mathbf{I}$ being the identity matrix. $\boldsymbol{\Omega}_{1}$ and $\boldsymbol{\Omega}_{2}$ are diagonal matrices, usually called the stiffness matrices, and $\mathbf{a},\mathbf{b},\mathbf{c}$ are constant vectors. By demanding that the interpolating function (\ref{eq14}) coincides with the theoretical solution at the endpoints of the interval $[t_{n},t_{n+1}]$, and that it satisfies the differential equation at $t=t_{n}$, we arrive at the recursion formula,
\begin{equation}
\boldsymbol{\Psi}_{n+1}=\boldsymbol{\Psi}_{n}+\mathbf{R}\mathbf{f}_{n}+\mathbf{S}\mathbf{f}_{n}^{(1)}, \label{eq15}
\end{equation}
where we use the notation $\mathbf{f}_{n}=\mathbf{f}(t_{n},\boldsymbol{\psi}_{n})$, $\mathbf{f}_{n}^{(1)}=\displaystyle{\left.\frac{\mathrm{d}}{\mathrm{d}t}\mathbf{f}(t,\boldsymbol{\Psi})\right|_{t=t_{n}}}$. $\mathbf{R}$ and $\mathbf{S}$ represent diagonal matrices defined as
\begin{equation}
\mathbf{R}=\boldsymbol{\Omega}_{2}\boldsymbol{\Phi}-\boldsymbol{\Omega}_{1}\boldsymbol{\varXi}, \qquad \mathbf{S}=\boldsymbol{\Phi}+\boldsymbol{\varXi}. \label{eq16}
\end{equation}
$\boldsymbol{\Phi}$ and $\boldsymbol{\varXi}$  are diagonal matrices with non-zero entries given by \cite{Fatunla1,Fatunla2},
\begin{eqnarray}
\Phi_{j}=\frac{e^{\Omega_{1,j}\delta t}-1}{\Omega_{1,j}\left(\Omega_{1,j}+\Omega_{2,j}\right)}, \label{eq17}
\end{eqnarray}
and
\begin{eqnarray}
\varXi_{j}=\frac{e^{-\Omega_{2,j}\delta t}-1}{\Omega_{2,j}\left(\Omega_{1,j}+\Omega_{2,j}\right)} .\label{eq18}
\end{eqnarray}
The recursion formula (\ref{eq15}) depends on the so far unknown stiffness matrices $\boldsymbol{\Omega}_{1}$ and $\boldsymbol{\Omega}_{2}$.  These matrices can be written in terms of the function $\mathbf{f}_{n}$ and its derivatives up to third orther in $t_{n}$.  The use of the Taylor expansion of $\Psi_{n+1}=\Psi(t_n+\delta t)$ and of the Maclaurin series of $\displaystyle{{e^{{\Omega _1}{\kern 1pt} \delta t}} = \sum\limits_{j = 0}^\infty  {\frac{{\delta {t^j}}}{{j!}}\,} {\Omega _{1,j}}}$
and of
$\displaystyle{{e^{ - {\Omega _2}{\kern 1pt} \delta t}} = \sum\limits_{j = 0}^\infty  {\frac{{\delta {t^j}}}{{j!}}{{\left( { - 1} \right)}^j}\,} {\Omega _{2,j}}}$, substituted in the recursion relation (\ref{eq15}), leads to a simple system of algebraic equations for $\Omega_1$ and $\Omega_2$. The components of the stiffness matrices obtained after solving these equations read as \cite{Fatunla2},

\begin{eqnarray}
\Omega_{1,j}&=&\frac{1}{2}\left(-D_{j}+\sqrt{D_{j}^{2}+4 E_{j}}\right), \nonumber \\
\Omega_{2,j}&=&\Omega_{1,j}+D_{j}, \label{eq19}
\end{eqnarray}
where $D_{j}$ and $E_{j}$ ($j=1,....,N$) are given in terms of the components of the derivatives $\mathbf{f}_{n}^{(k)}$  ($k=0,1,2,3$), 
\begin{eqnarray}
D_{j}&=&\frac{f_{n,j}^{(0)}f_{n,j}^{(3)}-f_{n,j}^{(1)}f_{n,j}^{(2)}}{f_{n,j}^{(1)}f_{n,j}^{(1)}-f_{n,j}^{(0)}f_{n,j}^{(2)}} ,\nonumber \\
E_{j}&=&\frac{f_{n,j}^{(1)}f_{n,j}^{(3)}-f_{n,j}^{(2)}f_{n,j}^{(2)}}{f_{n,j}^{(1)}f_{n,j}^{(1)}-f_{n,j}^{(0)}f_{n,j}^{(2)}}, \label{eq20}
\end{eqnarray}
provided that the denominator in Eq.(\ref{eq20}) is not zero. \\

Fatunla \cite{Fatunla2} has established that his method is L-stable and exponentially fitted to any complex value $\lambda$. This means that the corresponding stability function $R(\lambda\delta t)=\exp(\lambda\delta t)$, gives the optimum stability properties. Furthermore, it can be shown that the $j$th component of the local truncation error at $t=t_{n+1}$ is given by
\begin{eqnarray}
T_{n+1,j}=\frac{\delta t^{5}}{5!}\left[ f_{n,j}^{(4)}+\left(\Omega_{2,j}^{3}-\Omega_{2,j}^{2}\Omega_{1,j}+\Omega_{2,j}\Omega_{1,j}^{2}-\Omega_{1,j}^{3}\right) f_{n,j}^{(1)}\right. \nonumber \\
\left.-\Omega_{1,j}\Omega_{2,j}\left(\Omega_{1,j}^{2}-\Omega_{1,j}\Omega_{2,j}+\Omega_{2,j}^{2}\right) f_{n,j}^{0}\right]+\mathcal{O}(\delta t^{6}) .\label{eq21}
\end{eqnarray}

\bigskip
The implementation of Eq.(\ref{eq15}) to calculate $\boldsymbol{\Psi}_{n+1}$ requires the calculation of the function $\mathbf{f}_{n}$ and its first derivatives
$\mathbf{f}_{n}^{(1)}$ at each value of $t_{n}$, and also the stiffness matrices $\boldsymbol{\Omega}_{1}$ and $\boldsymbol{\Omega}_{2}$ to obtain
the matrices $\mathbf{R}$ and $\mathbf{S}$. We also calculate the truncation error $\mathbf{T}_{n+1}$ to control the size of the integration step
imposing a boundary criterion for $\left|\mathbf{T}_{n+1}\right|$. Note that to calculate the truncation error, we also need to evaluate 
$\mathbf{f}_{n}^{(4)}$.
\begin{figure}[h]
\begin{center}
\includegraphics[width=11cm,height=8cm]{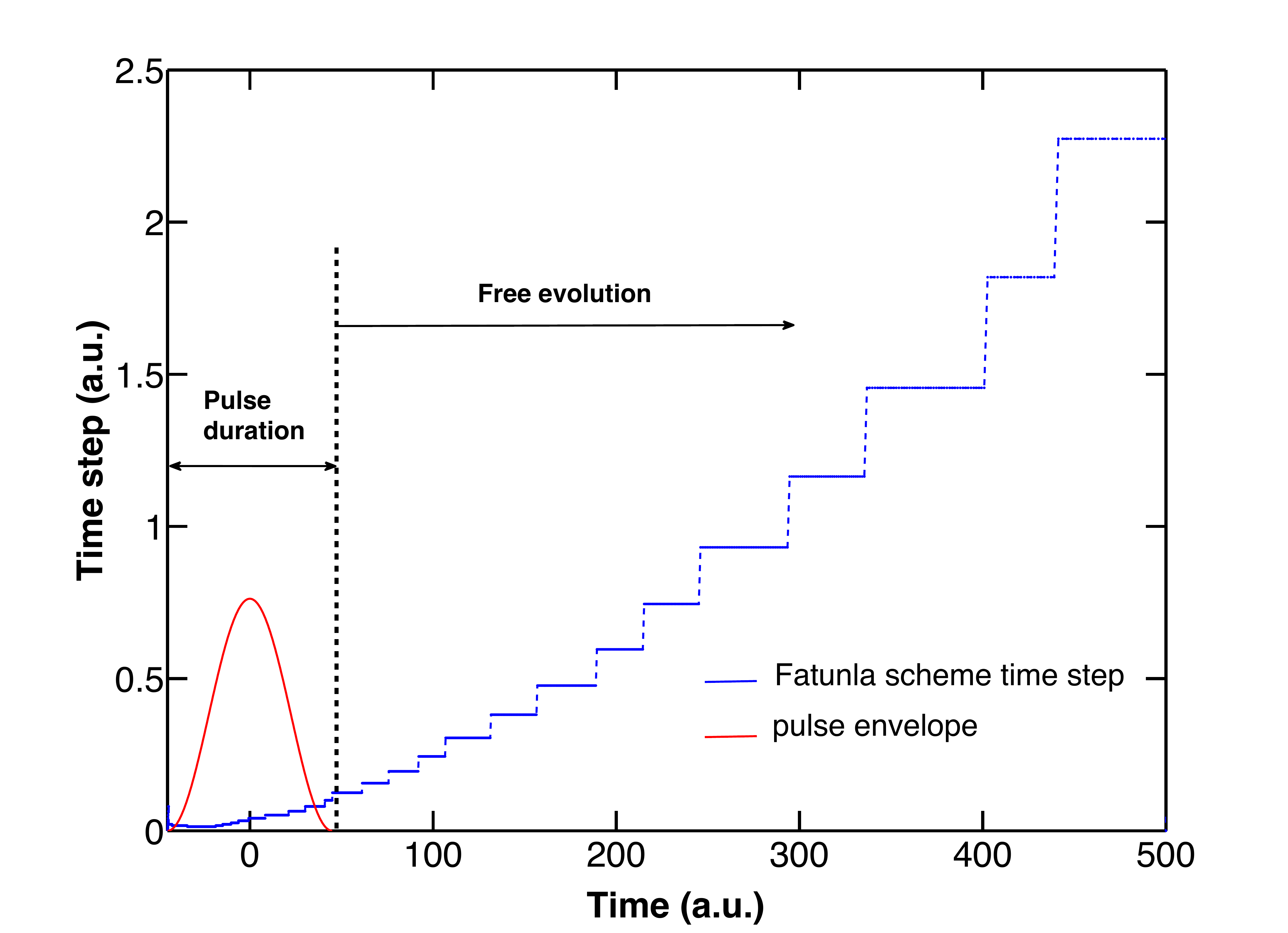}
\end{center}
\caption{(Color online) Evolution of the time step in Fatunla's method (blue line) for our Gaussian model problem. The cosine square pulse envelope (red line)  is also shown on an arbitrary scale.
The Gaussian potential parameters are $V_{0}=1$ a.u. and $\beta=1$ a.u. and we use an electromagnetic pulse with $I=10^{14}$ Watt/cm$^2$ peak intensity, $\omega=0.7$ a.u. photon energy and 
a duration of 10 optical cycles.}
\label{fig1}
\end{figure}
The stiffness parameters carry the intrinsic information on the natural oscillations of the system. Due to this fact, Fatunla's scheme can afford larger values of the time step compared with other explicit methods of Runge-Kutta type \cite{Madronero}.\\

\begin{figure}[b!]
\begin{center}
\includegraphics[width=11cm,height=8cm]{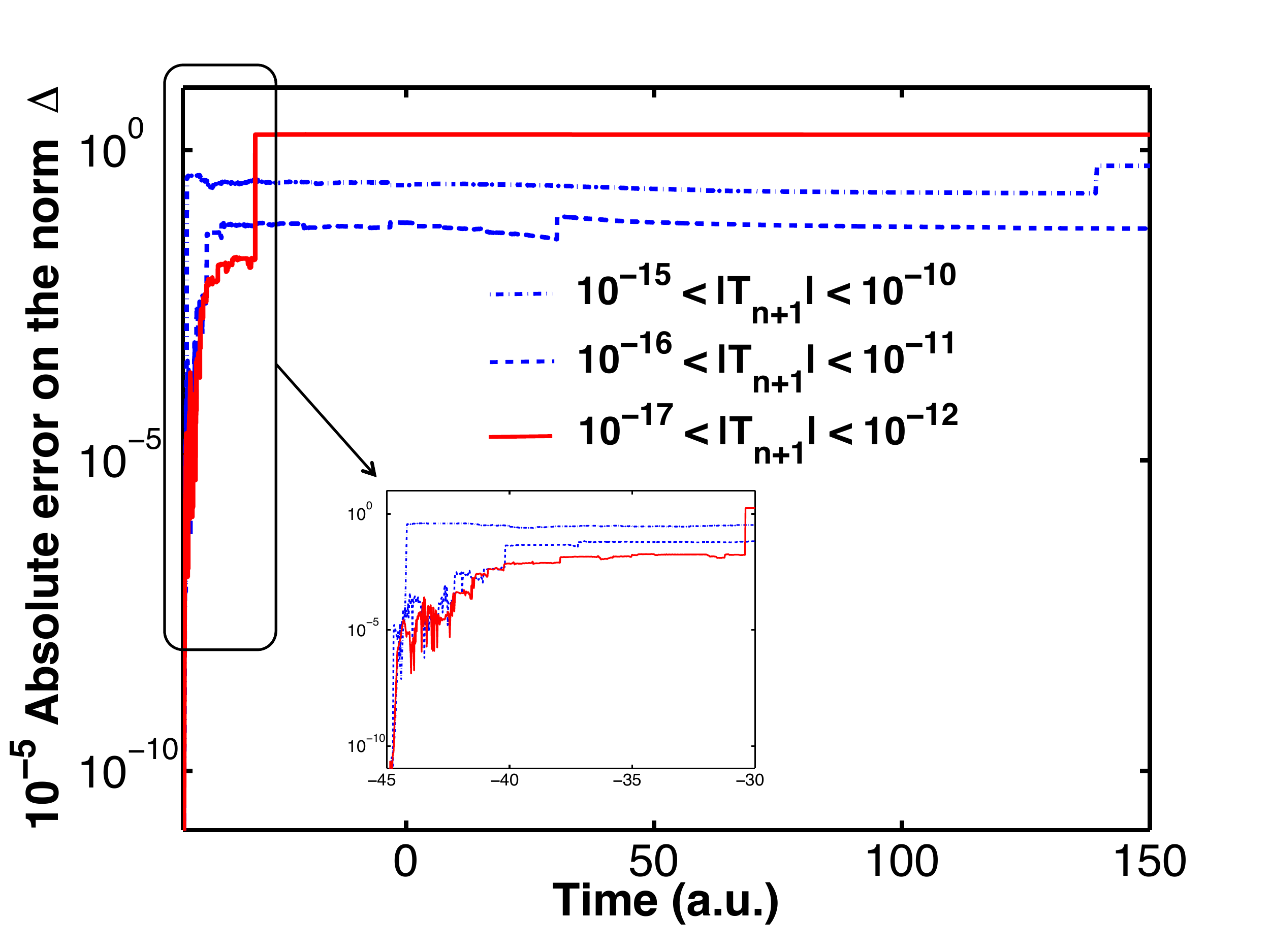}
\end{center}
\caption{(Color online) Absolute error in the norm $\Delta=\left\vert(\vert\Psi(\mathbf{r},t)|-|\Psi(\mathbf{r},t_{0})|)\right|$ on a logarithmic scale for different lower and upper bounds of the truncation error. The parameters of the Gaussian model problem are the same as in Fig.\ref{fig1}. The inset is a blow-up of the region at the beginning of the time propagation.}
\label{fig2}
\end{figure}

In Fig.\ref{fig1}, we show the evolution of the time step in Fatunla's propagation for our
Gaussian model problem. The pulse envelope is also plotted in arbitrary units to illustrate the duration of the pulse. We see that the 
time step becomes increasingly large after the end of the pulse, reaching values of around $2$ at the end of the total propagation ($500$ a.u. of time).
It is clear that the most demanding part of the propagation, and therefore the most time consuming one, is during the interaction of the pulse with the system.  This observation is important when it is necessary to propagate the wavefunction up to large distances after the end of the pulse, as is the case when the TSC method is used.
In the results for the time step shown in Fig.\ref{fig1}, the latter one is adapted according to the condition $10^{-14}<\lvert \mathbf{T}_{n+1} \rvert < 10^{-9}$, that is,
if the truncation error is lower than the lower bound $10^{-14}$ then we increase the time step, and if it is higher than the upper bound $10^{-9}$ it is
decreased. With this choice, the overall conservation of the norm is about $10^{-5}$, which is enough for the model problem case we are studying. 
For many physical problems, this level of accuracy in the norm is sufficient but, if a higher accuracy is needed, then we might expect that it is sufficient  to shift the bounds of the truncation error.  However, as shown in Fig.\ref{fig2}, such a conclusion is not correct. In Fig.\ref{fig2}, we consider three different constraints on the truncated error and calculate on a logarithmic scale, the absolute error on the norm denoted by $\Delta$ as a function of time. This error is defined as the absolute value of the difference between 1 and the norm at time t . In these three cases, the time propagation is started with the same time step namely $10^{-3}$ a.u. This time step always increases  while the truncated error is smaller than the prescribed lower bound and decreases if the truncated error is above the upper bound. In all three cases, we observe a significant loss of accuracy in $\Delta$ at the very beginning of the time propagation. As described by Madro\~nero and Piraux  \cite{Madronero}, this is due to initially very small values of the denominators in Eq. (20) which leads to inaccurate values of the stiffness matrix elements and of the truncated error. This problem is therefore intrinsically related to Fatunla's method and leads to difficulties in correctly controlling the time step. In fact , if we keep the time step constant from the beginning, we have a much better control of $\Delta$. We have also checked that this is true even in the field free case. On the other hand, in general, we see from the inset in Fig.\ref{fig2} that we maintain a higher accuracy when the constraint on the truncated error is more severe. In addition, we also observe several small jumps in $\Delta$ the magnitude of which are much smaller than the jump in $\Delta$ at the beginning of the time propagation. We attribute these jumps to an accumulation of roundoff errors. Indeed, we expect more roundoff errors in the case the constraint on the truncated error is the strongest since a smaller time step leads to a larger amount of calculations. Note that the jump observed in the red continuous line corresponds to a change of only one digit in the accuracy of the norm. The overall relative accuracy we obtain even for the most severe constraint we use on the truncation error is of the order of $10^{-5}$. To achieve a greater accuracy,  it is necessary to use a fixed and very small time step. These results show clearly that the achievable accuracy for the adaptive time step approach in Fatunla's method has a lower bound for a given initial time step.  As a result, the use of Fatunla's method rests on a compromise between the computer time required and the accuracy needed.  In the following, we consider the interaction of helium with a strong laser pulse. In that case, the accuracy on the norm reduces to about 4 significant digits when Fatunla's method is used. This prevents us to calculate the  probability of single ionization in various channels where the latter one is less than $10^{-4}$ a.u.  for field intensities currently used in the experiments.\\

In conclusion Fatunla's method allows one to treat stiff problems while fully exploiting the advantages of explicit schemes, namely that it only involves matrix vector multiplications.  However it has its own limitations.  
 
\subsection{Krylov subspace method}
\label{arnoldi}
In this section we consider a powerful method to propagate the TDSE solution, which provides accuracy of solutions and stability of propagation. It uses projection techniques on Krylov subspaces \cite{Saad}. This approach was proposed by Arnoldi \cite{Arnoldi} in the
calculation of the eigenstates of a matrix. Here we briefly recall the method used by Arnoldi
as a time propagator \cite{Park}, to solve the differential equation (\ref{eq5}). 
Since the overlap matrix is positive-definite, we can use the Cholesky decomposition $\mathbf{B}=\mathbf{U}^{\dagger}\mathbf{U}$
to form an orthonormal basis defining the new coefficients $\mathbf{\Phi}=\mathbf{U}\boldsymbol{\Psi}$. The TDSE for these coefficients is written in the form,
\begin{equation}
\frac{\mathrm{d}\mathbf{\Phi}(t)}{\mathrm{d}t}=-\mathrm{i} \mathbf{\widehat{H}}(t)\mathbf{\Phi}(t) ,\label{eq22}
\end{equation}
where $\mathbf{\widehat{H}}=(\mathbf{U}^{\dagger})^{-1}\mathbf{H} \mathbf{U}^{-1}$.
If we assume that the time interval is sufficiently small that the Hamiltonian may be treated as constant in time
over a time step $\delta t$, it is trivial to demonstrate that Eq.(\ref{eq22}) has a solution given by 
\begin{equation}
\mathbf{\Phi}(t+\delta t)=e^{-\mathrm{i} \mathbf{\widehat{H}}(t)\delta t} \mathbf{\Phi}(t). \label{eq23}
\end{equation}
If $\mathbf{\widehat{H}}$ is diagonalizable and can be written as $\mathbf{\widehat{H}}=\mathbf{E}\mathbf{\Lambda}\mathbf{E}^{-1}$, where
$\mathbf{\Lambda}$ is a diagonal matrix with the eigenvalues $\lambda_{i}$ of $\mathbf{\widehat{H}}$ on the main diagonal and 
$\mathbf{E}$ is the matrix with the corresponding eigenvectors of $\mathbf{\widehat{H}}$ as its columns, then Eq.(\ref{eq23})
can be reexpressed as follows
\begin{equation}
\mathbf{\Phi}(t+\delta t)=\mathbf{E} e^{-\mathrm{i}\mathbf{\Lambda}(t)\delta t} \mathbf{E}^{-1} \mathbf{\Phi}(t) .\label{eq24}
\end{equation}
However, for very large $N$ this may be unnecessary and computationally very demanding. Instead, we can  define the 
exponential in Eq.(\ref{eq23})  using a Taylor expansion of the form,
\begin{equation}
\mathbf{\Phi}(t+\delta t)=\left(\mathbf{I}-\mathrm{i} \delta t  \mathbf{\widehat{H}}(t)+... + \frac{(-\mathrm{i} \delta t)^{k}}{k!}  
\mathbf{\widehat{H}}^{k}(t)+...\right)\mathbf{\Phi}(t). \label{eq25}
\end{equation}
We use the successive matrix products as a basis set forming a Krylov subspace spanned by $(m+1)$ linearly independent vectors, denoted by
\begin{equation}
K_{m+1}=span\{\mathbf{\Phi},\mathbf{\widehat{H}}\mathbf{\Phi},...,\mathbf{\widehat{H}}^{m}\mathbf{\Phi}\}. \label{eq26}
\end{equation}
To build the Krylov subspace, we first use Gram-Schmidt orthogonalization of the initial vectors $\{\mathbf{\Phi},\mathbf{\widehat{H}}\mathbf{\Phi},...,\mathbf{\widehat{H}}^{m}\mathbf{\Phi}\},$
to obtain  an orthonormal basis $\{\mathbf{q}_{0},\mathbf{q}_{1},...,\mathbf{q}_{m}\}$. The procedure starts with $\mathbf{q}_{0}=\mathbf{\Phi}/\vert \mathbf{\Phi}\vert$, where the norm is defined as
$\vert \mathbf{\Phi}\vert= \sqrt{\mathbf{\Phi}^{\dagger}\cdot\mathbf{\Phi}}$. The $\mathbf{q}_{k}$ are obtained by calculating $\mathbf{\widehat{H}}\mathbf{q}_{k-1}$ and then orthonormalizing each vector with 
respect to $\mathbf{q}_{0},...,\mathbf{q}_{k-1}$. If we define $\mathbf{Q}$ to be a matrix formed by the $m+1$ column vectors $(\mathbf{q}_{0},...,\mathbf{q}_{m})$, we finally get 
\begin{equation}
\mathbf{\widehat{H}} \mathbf{Q}=\mathbf{Q} \mathbf{h},\label{eq27}                           
\end{equation}
giving
\begin{equation}
\mathbf{h}=\mathbf{Q}^{\dagger}\mathbf{\widehat{H}}\mathbf{Q}. \label{eq28}
\end{equation}
We see here that $\mathbf{h}$ is the Krylov subspace representation of the full Hamiltonian $\mathbf{\widehat{H}}$, and that in this procedure, we obtain simultaneoulsy the Krylov vectors 
$\mathbf{q}_{0},...,\mathbf{q}_{m}$. Arnoldi's algorithm is general and applies to non-hermitian matrices. It reduces the dense matrix $\mathbf{h}$ to an upper Hessenberg form, and in the particular case of hermitian matrices, to a
symmetric tridiagonal form. In this latter case, Lanczos has shown that this matrix can be obtained by means of a simple recursion formula. However, this formula is known to be problematic when the size of the Krylov subspace is large because the orthogonality of the Krylov vectors is rapidly lost \cite{Saad}. It is the reason why we do not use this algorithm in the present case.
Once we obtain the orthonormal Krylov subspace $\mathbf{Q}$ and the representation $\mathbf{h}$ of the Hamiltonian, it can be easily shown that Eq.(\ref{eq23}) can be written as
\begin{equation}
\mathbf{\Phi}(t+\delta t)=\mathbf{Q} e^{-\mathrm{i}\mathbf{h}\delta t} \mathbf{Q}^{\dagger} \mathbf{\Phi}(t). \label{eq29}
\end{equation}
The matrix $\mathbf{h}$ for all our case studies is tridiagonal, and its size is never bigger than $100 \times 100$, so the calculation of the exponential through direct diagonalization,
as in Eq.(\ref{eq24}), is straightforward.\\

In actual numerical calculations,  the Arnoldi algorithm \cite{Saad} requires some modifications. After a first calculation of a new Krylov vector 
$\mathbf{q}_{j+1}$, we ensure that the norm is equal to one, by re-checking the orthogonality against the previously calculated vectors, and perform again the Gram-Schmidt
procedure if necessary. In principle the orthogonality condition determines the maximum size of the Krylov subspace and the algorithm can be used with a number $m-1$ of vectors. Also, if we start generating the Krylov vectors from the ground state of the system, then $\mathbf{\Phi}(t=t_{0})$ is an eigenstate of the Hamiltonian, making it impossible to build a linearly independent set of Krylov vectors. To solve this problem, instead of using the vector $\mathbf{\Phi}(t=t_{0})$ as a starting point, we use a modified vector $\mathbf{\Phi}(t=t_{0})+\boldsymbol{\varrho}$, with $\boldsymbol{\varrho}$ a vector of random entries no larger than $10^{-10}$.\\
\begin{figure}[ttp]
\begin{center}
\includegraphics[width=11cm,height=9cm]{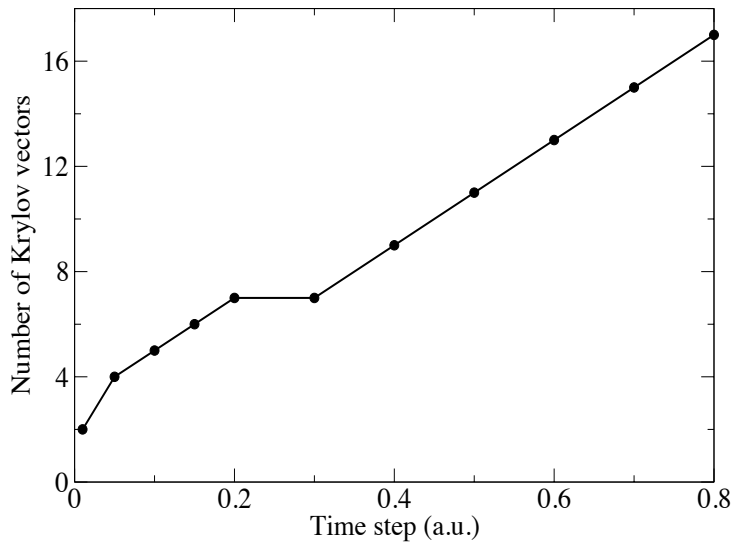}
\end{center}
\caption{Number of Krylov vectors required to obtain convergence of the final vector propagated for different values of the (fixed) time step.
The parameters of the Gaussian model problem are the same as in Fig.\ref{fig1}.}
\label{fig3}
\end{figure}

By construction (see Eq.\ref{eq23}), the stability function associated to the numerical time propagator based on Arnoldi's algorithm is given by $R(\lambda\delta t)=\exp(\lambda\delta t)$. As a result, it has exactly the same stability properties as Fatunla's algorithm. However, it is worth remembering that Eq.(\ref{eq23}) is only valid if the Hamiltonian is time independent. It is therefore a good approximation only for small values of $\delta t$. In the present case, there are two types of errors. The first one is directly related to Arnoldi's algorithm for the calculation of the exponential of a matrix. This type of error has been discussed in detail  by Saad \cite{Saad2} and later on by Hochbruck and Lubich \cite{Hochbruck}. We have checked that this type of error is always negligible and does not depend on the time step. The second type of error is due to assuming that the Hamiltonian does not depend explicitly on time over the time step $\delta t$. We  estimated this type of error by calculating $\|\mathrm{d}\mathbf{\Psi}/\mathrm{d}t+\mathrm{i}\mathbf{H}\mathbf{\Psi}\|$ and checked that, as expected, it is of the order $\delta t^2$. Another way to estimate  this type of error is to compare our results with those obtained with an Arnoldi based method that takes explicitly into account the time dependence of the Hamiltonian. This can be done by using a Magnus expansion  of the time evolution operator \cite{Magnus,Iserles}. However, this method requires very time consuming calculations beyond the scope of this contribution.  On the other hand, as it was already noted by other authors \cite{Park}, enlarging the size of the Krylov space allows for larger time steps to be considered. In Fig.\ref{fig3}, we give the number of Krylov vectors necessary to obtain convergence of our results as a function of the time step used in the calculations for the case of the one-dimensional Gaussian model potential with the same parameters  as in Fig.\ref{fig1}. In our calculations, the time step $\delta t$ is kept constant during the propagation. The choice of the optimal value of the time step and of the corresponding dimension of the Krylov space is therefore the result of a compromise while trying to reduce the computer time.\\

The innovative use of the Arnoldi method as an explicit approach offers then the convenience that we only require matrix-vector and scalar products, which then transforms the method in a time-efficient approach as is the case for Fatunla's method. Furthermore, this particular scheme is norm-conserving with the advantage of providing a check for the method, even though it also means that it is not easy to quantify the error in the calculation of the norm. In the following sections the accuracy of both methods is tested in various situations by using a high order predictor-corrector method.

\subsection{A predictor-corrector method.}
\label{predcor}

 In this subsection, we briefly describe the predictor-corrector (P-C) scheme we use to test the accuracy of both explicit methods described above. The predictor is either Fatunla's or Arnoldi's algorithm. The corrector is a fully implicit method of Runge-Kutta type which, here, is a four-stage Radau IIA method of order 7. \\
 
In a general Runge-Kutta method,  the numerical solution $\mathbf{\Psi}_{n+1}$ of Eq.(\ref{eq8}) at a given time $t=t_{n+1}$ is obtained from the solution $\mathbf{\Psi}_n$ at time $t=t_n$ as
\begin{equation}
\boldsymbol{\Psi}_{n+1}=\boldsymbol{\Psi}_{n}+\delta t \sum_{i=1}^{s}b_{i}f(t_{i},\mathbf{Y}_{i}), \label{eq30}
\end{equation}
where $\delta t$ is the time step and $f(t_{i},\mathbf{Y}_{i})=-\mathrm{i} \mathbf{B}^{-1}\mathbf{H}(t_{i})\mathbf{Y}_{i}$ with $t_{i}=t_{n}+c_{i}\delta t$. $b_{i}$ and
$c_{i}$ are coefficients defining the Runge-Kutta method for a number $s$ of stages. We assume that the solution vector $\mathbf{\Psi}$ is of dimension $N$. The quantities $\mathbf{Y}_{i}$ estimate the solution $\mathbf{\Psi}$ at the intermediate time $t_i$. They are obtained  by solving the following linear ($sN\times sN$) system of equations
\begin{equation}
\mathbf{Y}_{i}=\boldsymbol{\Psi}_{n}+\delta t \sum_{k=1}^{s}a_{ik}f(t_{k},\mathbf{Y}_{k}), \label{eq31}
\end{equation}
where the $a_{ik}$ are again given by the method. Solving such system represents the main difficulty of an implicit Runge-Kutta scheme. If this scheme is used for the corrector, we could in principle avoid solving such system of equations by using an iterative procedure in which we replace the vector $\mathbf{Y}_k$ in the right hand side of Eq.(\ref{eq31}) by $\mathbf{Y}^{(j-1)}_k$ where $j$ gives the order of the iteration process. At the order $0$, $\mathbf{Y}^{(0)}_k$ is provided by the predictor. However, such an iterative procedure is not stable. Instead, we follow a different iterative  procedure that has been developed by van der Houwen and Sommeijer \cite{Houwen1,Houwen2}. By introducing a diagonal matrix whose entries are calculated to guarantee optimum stability properties, they transform the ($sN\times sN$) system (\ref{eq31}) into a set of uncoupled ($N\times N$) systems of equations that can be solved in parallel at each iteration. More precisely, they rewrite Eq.(\ref{eq31}) as follows
\begin{equation}
\mathbf{Y}_{i}^{(j)}-\delta t \, d_{ii} \, f(t_{i},\mathbf{Y}_{i}^{(j)})=\boldsymbol{\Psi}_{n}+
\delta t \sum_{k=1}^{s}(a_{ik}-d_{ik})f(t_{k},\mathbf{Y}_{k}^{(j-1)}), \label{eq32}
\end{equation}
where the $d_{ik}$ are the entries of the diagonal matrix. The iterations in Eq.(\ref{eq32}) start with $\mathbf{Y}_{i}^{(0)}$, provided  by the predictor. The iteration scheme is performed
until a value $j=$\textbf{max}$_{cor}$ for which we have convergence.  Then we can replace $\mathbf{Y}_{i}=\mathbf{Y}_{i}^{(m)}$ in Eq. (\ref{eq30}) to obtain the solution at $t=t_{n+1}$. Once we have calculated $\mathbf{\Psi}_{n+1}$, we can evaluate its norm and use its conservation as a criterion to monitor the size of the time step.\\

\begin{figure}[h]
\begin{center}
\includegraphics[width=11cm,height=8cm]{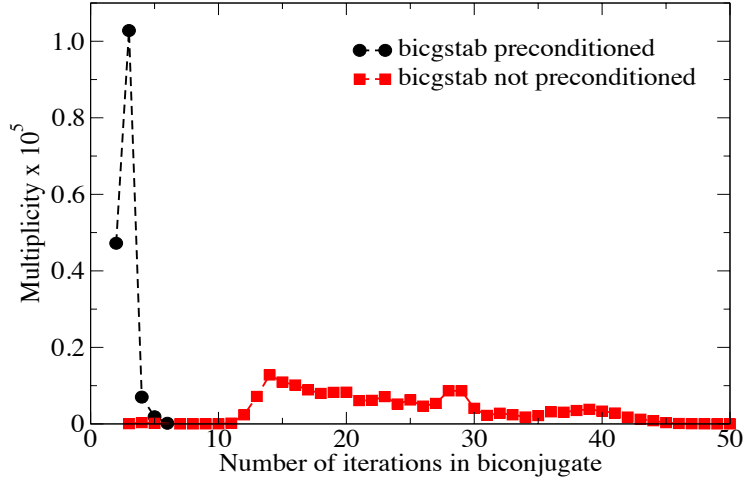}
\end{center}
\caption{(Color online) Number of iterations in the Bi-CGSTAB and their multiplicities during the propagation with the electromagnetic pulse. The parameters
of the Gaussian model problem are $V_{0}=4$ a.u., $\beta=0.1$ a.u., with a pulse of peak intensity $I=10^{16}$ Watt/cm$^2$, photon energy $\omega=0.5$ a.u. and duration of 8 optical cyles.}
\label{fig4}
\end{figure}
\begin{figure}[b!]
\begin{center}
\includegraphics[width=11cm,height=8cm]{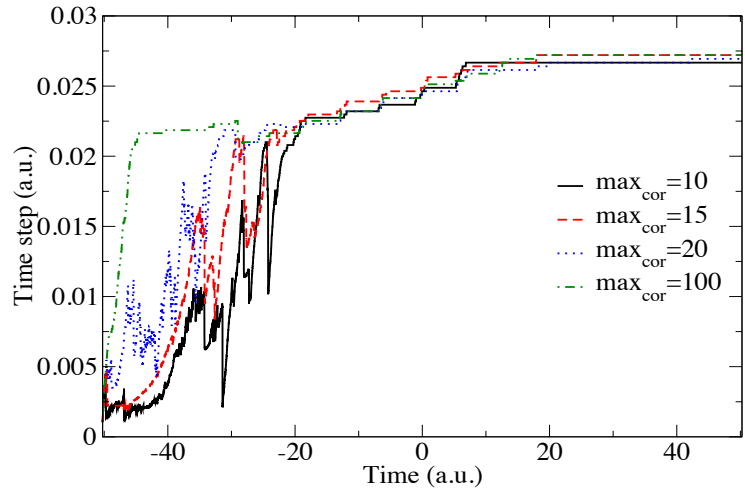}
\end{center}
\caption{(Color online) Time step evolution for P-C method of time propagation during the propagation with the electromagnetic pulse. The parameters
of the Gaussian model problem are the same as in Fig.\ref{fig4}.}
\label{fig5}
\end{figure}
Using the P-C method requires  solving a large number of $(N\times N)$ systems of equations. To solve these systems, we use an iterative method known as the biconjugate gradient stabilized method (Bi-CGSTAB) \cite{Vorst}. In order to reduce drastically the number of iterations, we use a pre-conditioner based on an incomplete LU factorization. In Fig.\ref{fig4}, we consider the case of our one-dimensional Gaussian model potential with $V_0=4$ a.u. and $\beta=0.1$ a.u.  and a pulse of frequency $\omega=0.5$ a.u., duration of 8 optical cycles and peak intensity $ I=10^{16}$ Watt/cm$^2$.  We show, in this Fig.\ref{fig4}, the multiplicity as a function of the number of iterations in the Bi-CGSTAB during the interaction. By multiplicity, we mean the number of times a given number of iterations is repeated during the whole propagation.
It can be seen that without pre-conditioner, the number of iterations can grow significantly before reaching convergence, while using the preconditioned Bi-CGSTAB, the number of iterations is  maintained below five. This reduces the computational time needed  by 25\%. However, care must be taken when including a pre-conditioner, since, by increasing the number of operations,
it may increase the computational times even though it accelerates convergence. 
As mentioned above, the corrector scheme is iterated up to  $j=$\textbf{max}$_{cor}$ where convergence is achieved.  In Fig.\ref{fig5} we plot the time evolution of the time step during the  propagation for different values of the maximum of iterations \textbf{max}$_{cor}$ in the corrector . We see here that, as we increase this maximum number
of iterations, the value of the time step becomes larger. The relative error in the norm which is the same for all the calculations is of the order of $10^{-11}$.
Moreover, the computational time with \textbf{max}$_{cor}$=100 is half the time consumed for \textbf{max}$_{cor}$=10 because it allows to use a much larger time step. It is therefore 
advisable to use large values of \textbf{max}$_{cor}$ to speed up the calculations.

\section{Results}

\subsection{Model potential}

In this section we first present results for our case study of the one-dimensional Gaussian potential taking $V_0=1$ and $\beta=1$.   The electron wavepacket is developed in a basis of 200 B-splines and we use the time scaled coordinate method during the propagation \cite{Hamido}.   
\begin{figure}[!h]
\begin{center}
\includegraphics[width=10cm,height=6.6cm]{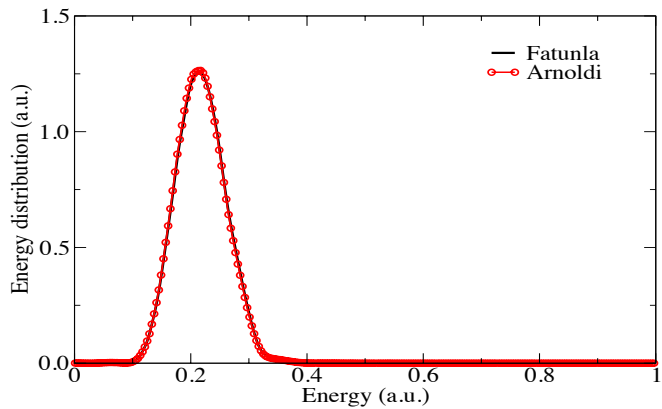}
\end{center}
\caption{(Color online) Energy distribution for the Gaussian model potential. The time propagation uses  Fatunla's propagator with adaptive time step and Arnoldi's propagator with five Krylov vectors and a fixed time step $\delta t = 0.3$ a.u. The parameters
of the model problem are $V_0=1$ and $\beta=1$  with a pulse of peak intensity I=$10^{14}$ Watt/cm$^2$, photon energy $\omega=0.7$ a.u. and a duration of $10$ optical cycles. The relative difference between both curves is of the order of $10^{-3}$.}
\label{fig6}
\end{figure}
We run our codes on an INTEL XEON 2.33 GHz Processor 51.40 (32 GB Ram). We  choose a pulse of frequency $\omega=0.7$ a.u. and a full duration of 90 a.u. of time which corresponds to 10 optical cycles. The peak intensity I = $10^{14}$ Watt/cm$^2$. 
\begin{figure}[!h]
\begin{center}
\includegraphics[width=9.5cm,height=6.8cm]{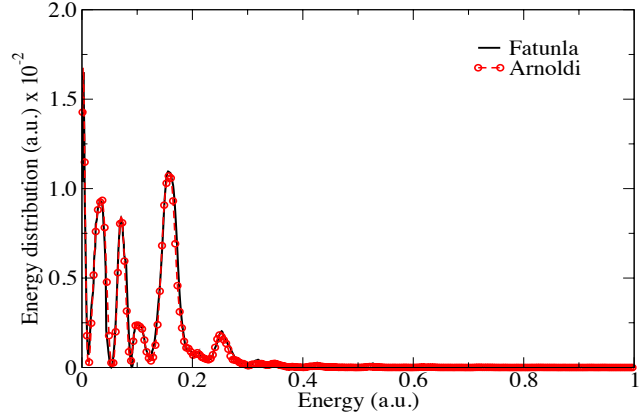}
\end{center}
\caption{(Color online) Energy distribution for the Gaussian model potential. The time propagation uses  Fatunla's propagator with adaptive time step and Arnoldi's propagator with $20$ Krylov vectors and a fixed time step $\delta t = 0.1$ a.u. The parameters
of the model problem are $V_0=1$ and $\beta=1$  with a pulse of peak intensity $I=10^{14}$ Watt/cm$^2$, photon energy $\omega=0.1$ a.u. and duration of $10$ optical cycles.The relative difference between both curves is of the order of $10^{-3}$.}
\label{fig7}
\end{figure}
\begin{figure}[!b]
\begin{center}
\includegraphics[width=12cm,height=10cm]{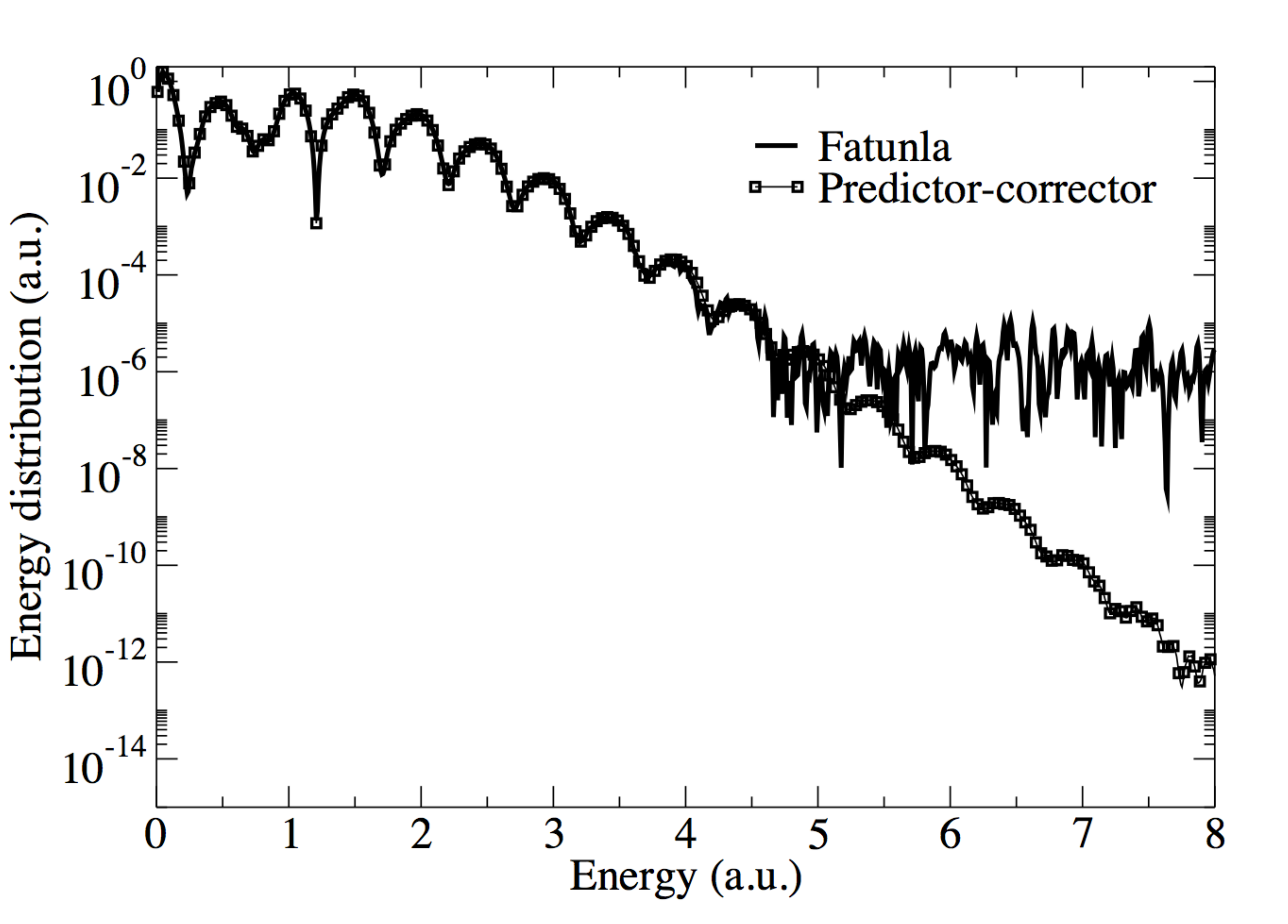}
\end{center}
\caption{Energy distribution for the Gaussian model potential. The time propagation uses  Fatunla's propagator with adaptive time step and the predictor-corrector method. The parameters
of the model problem are $V_0=4.0$ and $\beta=0.1$  with a pulse of peak intensity $I=10^{16}$ Watt/cm$^2$, photon energy $\omega=0.5$ a.u. and duration of $8$ optical cycles.}
\label{fig8}
\end{figure}
In Fig.\ref{fig6},  the energy distribution is calculated by propagating the scaled wavepacket to a stationary state until a time of 1500 a.u. when convergence is achieved.  The results shown are obtained using the two explicit propagators.  Both methods converge to the same result but Fatunla's propagator uses 2.3 s of computer time with an adaptive time step while Arnoldi's propagator using five Krylov vectors and a fixed time step $\delta t = 0.3$ a.u. takes 6.2 s. For these values of intensity and frequency both methods give easily the correct result.  However Arnoldi's method performs poorly from a computer time point of view.  This can be understood by referring to Fig.\ref{fig1} where we show that Fatunla's propagator allows the use of ever larger time steps, particularly during the propagation after the end of the pulse, while Arnoldi's propagator keeps the same time-step throughout the propagation.\\
 
\begin{figure}[ttp]
\begin{center}
\includegraphics[width=12cm,height=10cm]{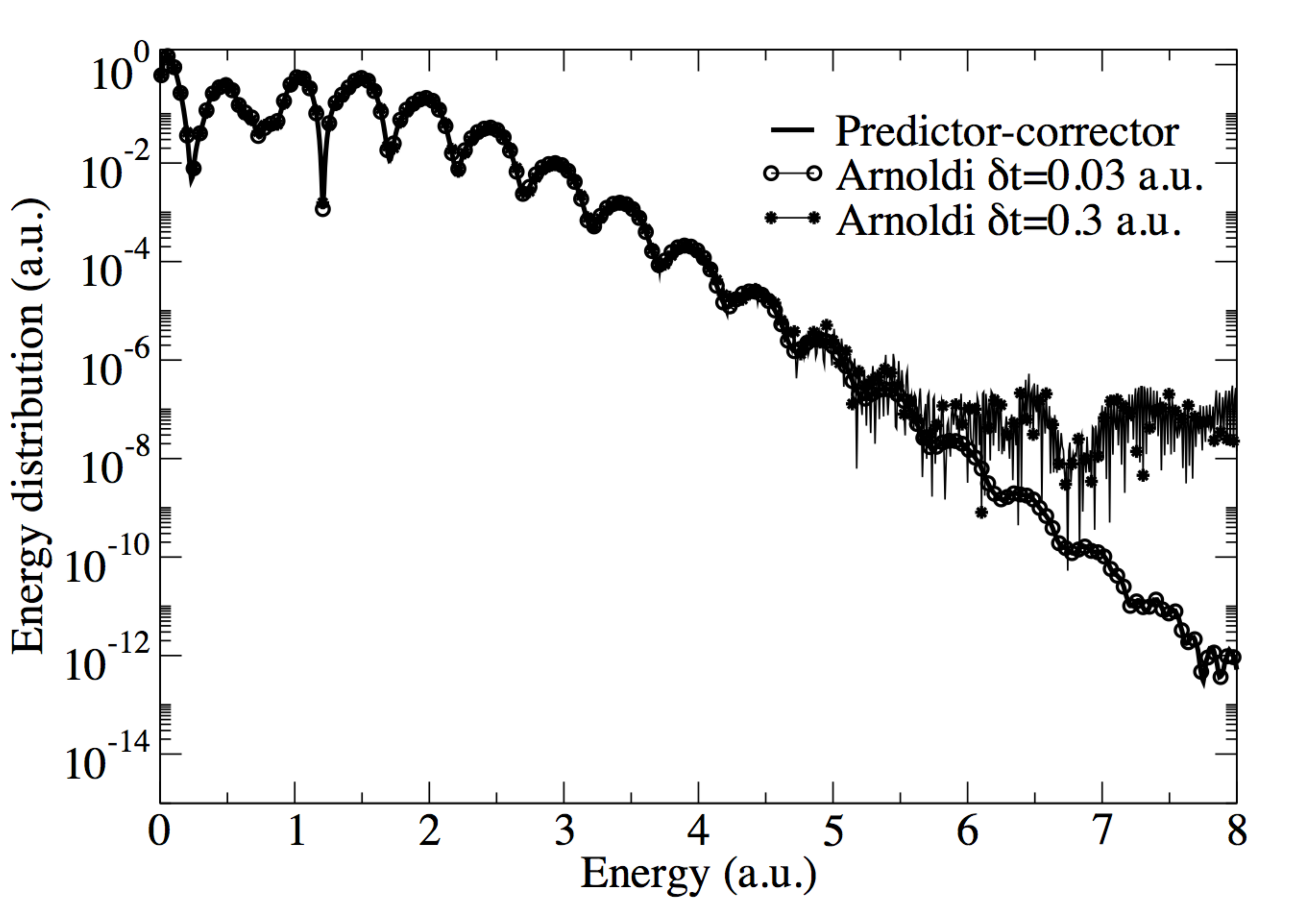}
\end{center}
\caption{Energy distribution for the Gaussian model potential. The time propagation uses  Arnoldi's propagator with fixed time step and the P-C method with adaptive time step. The parameters
of the Gaussian model problem are as in Fig.\ref{fig8}.}
\label{fig9}
\end{figure}

To check how these methods behave in a more challenging case, we consider the same model potential with a pulse of frequency 0.1 a.u. with the same number of optical cycles and peak intensity.  In this case, the total pulse duration is equal to 630 a.u. We see  in Fig.\ref{fig7} that both methods give identical results.  These results are obtained after propagating the wavepacket up to a time of 2500 a.u.  The running time with Fatunla's propagator is equal to 958.27 s while in this case, Arnoldi's propagator performs better using 553.69 s for a subspace of 20 Krylov vectors and a fixed time step $\delta t = 0.1$ a.u.  It can be seen that in general, Arnoldi's propagator performs better than Fatunla's propagator for long pulses.\\

To further probe these methods we increase the number of bound states supported by our potential by choosing  ${V_0} = 4$ and $\beta=0.1$.  The pulse has a frequency $\omega=0.5 $ a.u. with a duration of 100.53 a.u. that corresponds to 8 optical cycles. The peak intensity I = $10^{16}$ Watt/cm$^{ - 2}$. In this case, we use 1700 B-splines to propagate the wavepacket up to a time of 5000 a.u. Fig.\ref{fig8} shows the energy distribution obtained using Fatunla's propagator (straight line) and the predictor-corrector scheme (squares), which is used to test the accuracy of Fatunla's method.  Comparison of these two methods shows that Fatunla keeps the accuracy in the results  down to a value of  $10^{-5}$ a.u. for the energy distribution.  The TSC approach is used with an asymptotic scaling factor of 0.1.  Fatunla's propagator takes 379.36 s while the P-C method with adaptive time step takes 1801.49 s.  It is clear that Fatunla uses remarkably less computer time and works as long as the accuracy required is up to six digits. Fig.\ref{fig9} shows the energy distribution obtained with Arnoldi's propagator for the same parameters as in Fig.\ref{fig8}.  We show the results obtained with Arnoldi's approach and two different fixed time steps and compare these results with those obtained with the P-C scheme. The Krylov subspace contains 20 vectors and the wavepacket is again propagated up to 5000 a.u.  The circles show the results for a time step $\delta t=0.03$ a.u. and the stars for $\delta t=0.3$ a.u.  We note that increasing the time step leads to less accurate results by comparison with the P-C method. Arnoldi scheme takes 5957.19 s for a time step of  0.03 a.u. and 598.11 s for a time step of 0.3 a.u.\\

\subsection{Hydrogen Atom}

\begin{figure}[!ht]
\begin{center}
\includegraphics[width=10cm,height=7cm]{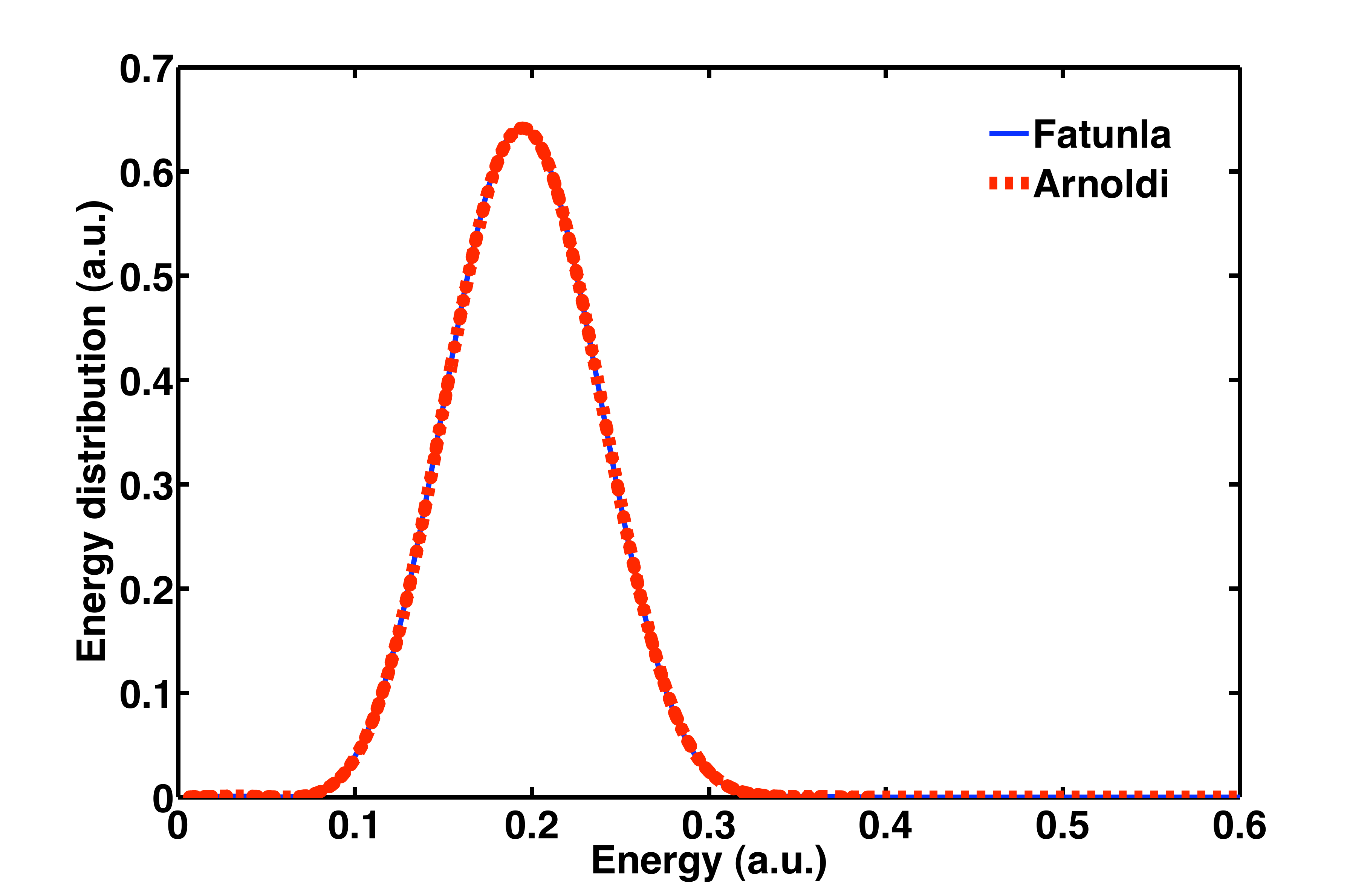}
\end{center}
\caption{(Color online) Energy distribution resulting from the interaction of the hydrogen atom with a cosine square pulse.  Fatunla's and Arnoldi's propagators are used. The pulse has a peak intensity I=10$^{14}$ Watt/cm$^2$, a frequency $\omega=0.7$ a.u. and a duration of 10 optical cycles. The basis-set of functions used is a set of 100 Coulomb Sturmian functions per angular momentum. Ten angular momenta are included and the non-linear parameter $\kappa$ of the Coulomb Sturmian functions is equal to 0.3.  The Arnoldi propagator uses 5 Krylov vectors and a time step of $\delta t = 0.05$ a.u.The relative difference between both curves is of the order of $10^{-3}$.}
\label{fig10}
\end{figure}

We now apply these methods to the more complex case of the interaction of hydrogen atom with a cosine square laser pulse.  We use a spectral method based on the expansion of the wavefunction in a basis of Coulomb Sturmian functions \cite{Madronero}, without implementing the TSC method.  Unless otherwise stated, we performed all calculations on a laptop (with an INTEL core 2 duo processor of 2.4 GHz). The first pulse we use has a frequency of 0.7 a.u., a duration of 10 optical cycles and an intensity I=10$^{14}$ Watt/cm$^2$, as in the case of Fig.\ref{fig6}.  In these rather simple conditions, we use 10 angular momenta.  The non-linear parameter $\kappa$ of the Coulomb Sturmian functions is taken equal to 0.3 a.u.  Fatunla's and Arnoldi's algorithms  produce the converged energy distribution as shown in Fig.\ref{fig10}.  The calculations carried on with  Fatunla's propagator and an adaptive time step  take 10.50 s of computer time. The integration performed with Arnoldi's method takes 13.72 s.  For a time step of $\delta t=0.05$ a.u. and 100 Coulomb Sturmian functions per angular momentum, it needs only 5 Krylov vectors.  In this case Arnoldi's method is slower than Fatunla's method.  This is related to the number of basis-set functions used.  As this number increases, higher eigenvalues are generated in the Hamiltonian spectrum thereby increasing the stiff character of the system of equations to solve. In that case, more Krylov vectors have to be included to maintain the accuracy of the results.
\begin{figure}[!ht]
\begin{center}
\includegraphics[width=10cm,height=6.8cm]{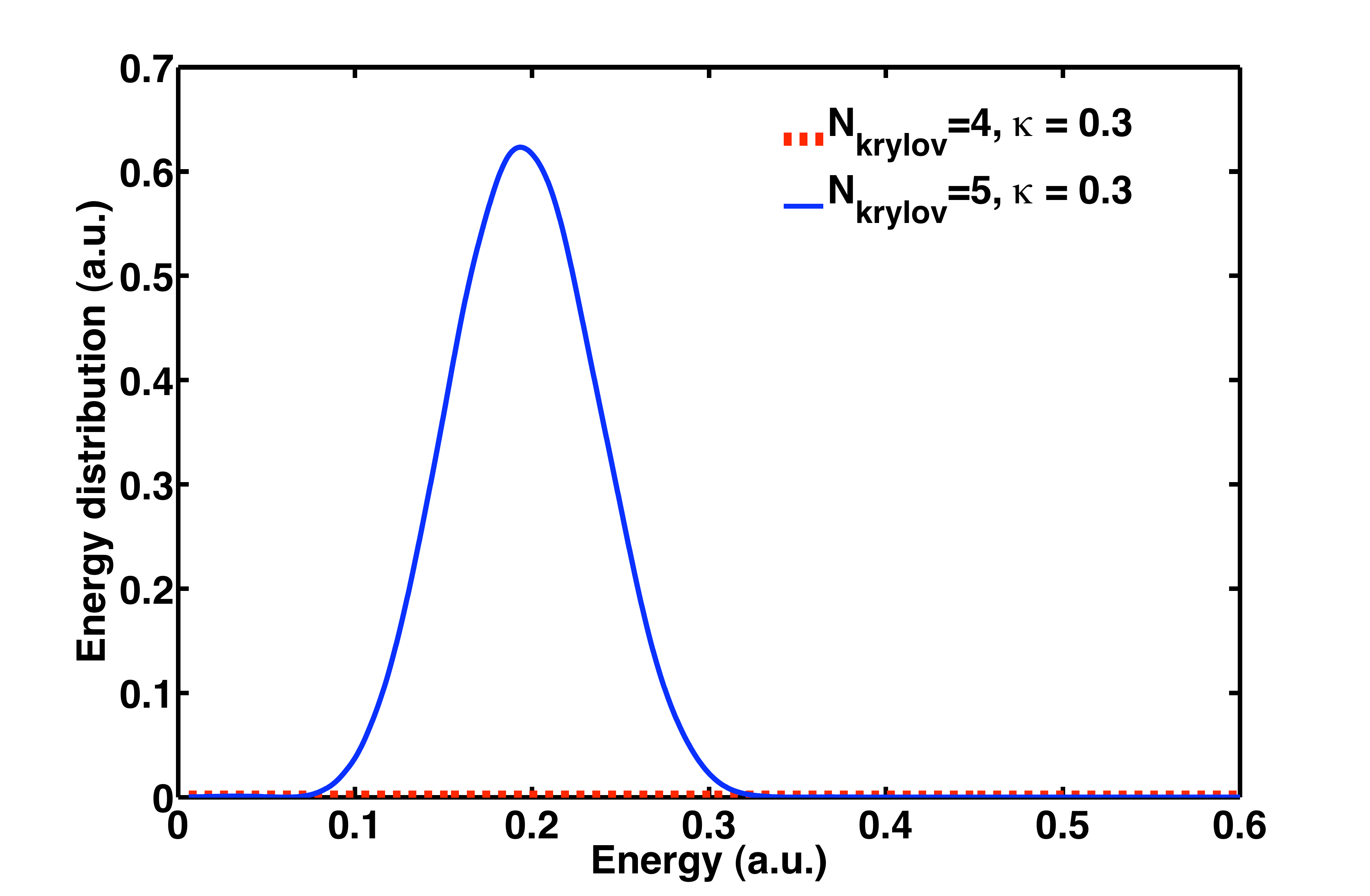}
\end{center}
\caption{(Color online) Energy distribution resulting from the interaction of the hydrogen atom with a cosine square pulse.  Arnoldi's propagator is used. The pulse parameters are as in Fig.\ref{fig10}, using 100 Coulomb Sturmian functions per angular momentum. Ten angular momenta are included and the non-linear  parameter $\kappa$ of the Coulomb Sturmian functions is equal to 0.3.  For a time step of $\delta t = 0.05$ a.u., we compare results when 5 and 4 Krylov vectors are used.}
\label{fig11}
\end{figure}

\begin{figure}[h]
\begin{center}
\includegraphics[width=10cm,height=7.2cm]{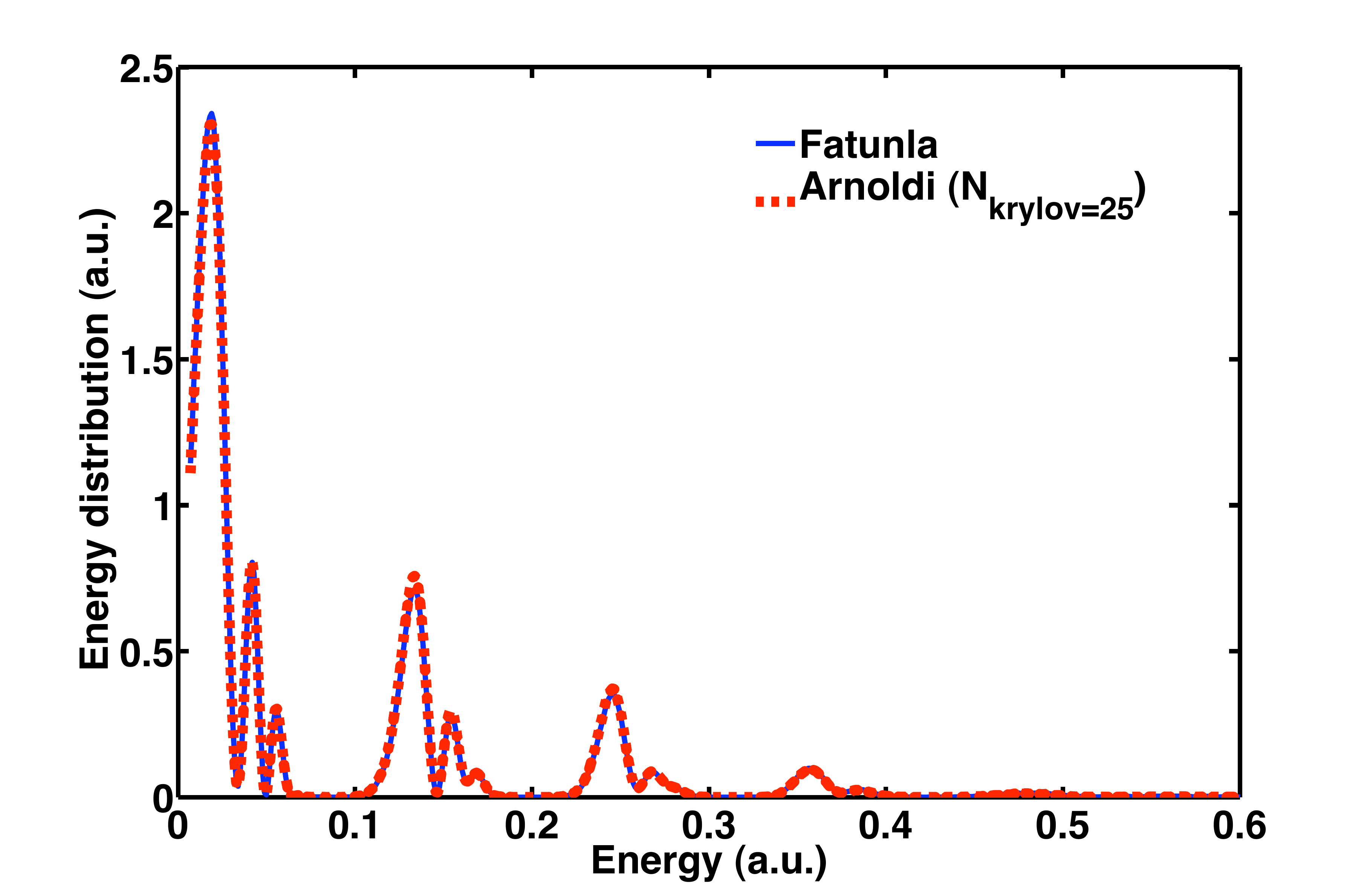}
\end{center}
\caption{(Color online) Energy distribution resulting from the interaction of the hydrogen atom with a cosine square pulse.  Arnoldi's propagator is used. The pulse has a peak intensity I=10$^{14}$ Watt/cm$^2$, a frequency $\omega=0.114$ a.u. and a duration of 20 optical cycles. We use a set of 600 Coulomb Sturmian functions per angular momentum. Ten angular momenta are taken into account and the non-linear parameter of the Coulomb Sturmian functions $\kappa=0.3$.   The Arnoldi propagator uses 25 Krylov vectors and a time step of $\delta t = 0.05$ a.u. The relative difference between both curves is of the order of $10^{-3}$.}
\label{fig14}
\end{figure}
\begin{figure}[!ht]
\begin{center}
\includegraphics[width=10cm,height=7.2cm]{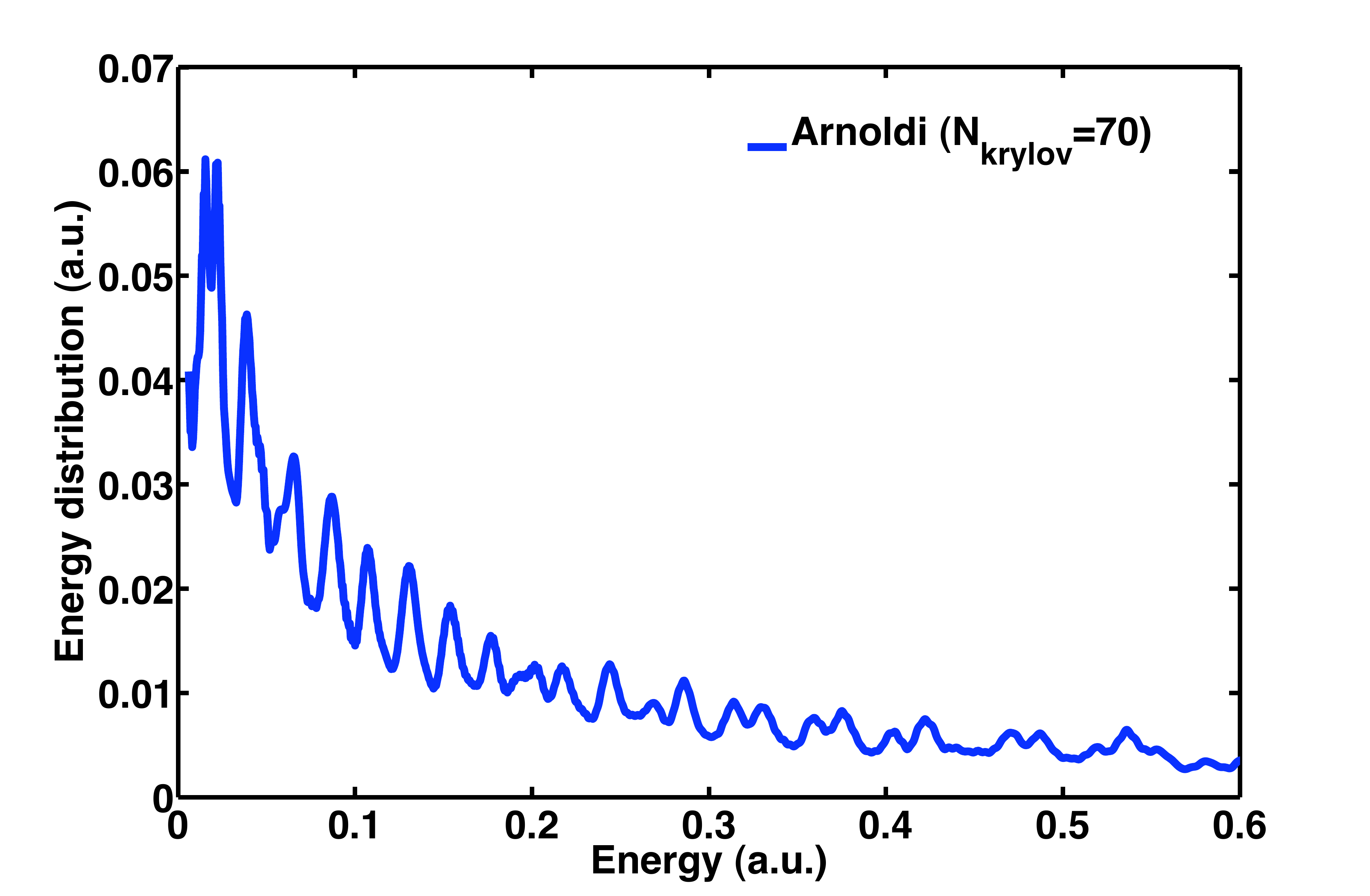}
\end{center}
\caption{(Color online)Energy distribution resulting from the interaction of the hydrogen atom with a cosine square pulse.  Arnoldi's propagator is used.  The pulse has a peak intensity I=10$^{14}$ Watt/cm$^2$, a frequency $\omega=0.0228$ a.u. and a duration of 4 optical cycles. The basis-set of functions used is a set of 1200 Coulomb Sturmian functions per angular momentum. 80 angular momenta are taken into account and the non-linear parameter $\kappa$ of the Coulomb Sturmian functions is equal to 0.3.  The Arnoldi propagator uses 70 Krylov vectors and a time step of $\delta t = 0.05$ a.u.}
\label{fig15}
\end{figure}

In Fig.\ref{fig11} we illustrate the effect of reducing the number of Krylov vectors $n_k$ from 5 to 4.  It is surprising to see that the propagator gives a completely flat spectrum when the dimension of the Krylov space is insufficient.  Fig.\ref{fig11} shows that for a basis set of 100 Coulomb Sturmian functions per angular momentum, accurate results for the energy distribution require  a minimum of 5 Krylov vectors.\\

These calculations performed in a Coulomb Sturmian basis can be further tested by varying their non-linear parameter $\kappa$. If instead of using $\kappa=0.3$, we use $\kappa=0.4$, all the other parameters remaining the same,  we again obtain a completely flat energy distribution.  By increasing the value of the non-linear parameter $\kappa$, the value of the eigenenergies increases thereby  increasing the stiff character of the problem.  To successfully reproduce an accurate energy distribution we now would need to increase the number of Krylov vectors. If on the other hand, we keep the value of $\kappa$ equal to 0.3 and  increase the number of basis functions, converged results  are only obtained when 8 Krylov vectors are used.  The increase in the number of  Coulomb Sturmians generates higher eigenenergies thereby increasing again the stiff character of the system. The eigenvalues of matrix \textbf{h} range from the eigenvalue of the initial state (by construction) to approximately the highest one of matrix \textbf{H}. In summary, any change which results in a higher maximum eigenvalue for \textbf{H} necessitates an increase in the number of Krylov subspace vectors required for convergence.

 It is interesting to note that to get an accurate spectrum, one of the eigenvalues of \textbf{h} must converge to 0.2 which corresponds to the position of the maximum of the spectrum which is what we expect from energy conservation ($0.2=-0.5+\omega$). If none of the eigenvalues converges to 0.2, the spectrum is completely flat because all the eigenvalues of \textbf{h} which are usually very high except the first one, do not contribute significantly to the spectrum. In addition, it is important to stress that decreasing the time step does not modify the minimum number of Krylov vectors to be used.\\

In Fig.\ref{fig14} we compare the performance of  Arnoldi's and Fatunla's methods for a more difficult case. We consider a pulse of  frequency $\omega=0.114$ a.u. and a duration of 20 optical cycles.  The pulse intensity is the same as before, I=10$^{14}$ Watt/cm$^2$. We use a basis set of  600 Coulomb Sturmian functions  per angular momentum.  10 angular momenta are included in the calculations and the non-linear parameter $\kappa=0.3$. Both energy distributions agree but Fatunla's scheme, which needs a very small time step, takes 66004 s of computer time while Arnoldi's method takes 1419 s with 25 Krylov vectors and a time step of 0.05 a.u.  This case illustrates clearly that  Arnoldi's algorithm copes  in an efficient way with the stiffness of the problem by increasing the size of the Krylov subspace.\\

In Fig.\ref{fig15} we show results obtained for the challenging case of a pulse of very low frequency $\omega=0.0228$ a.u. and a duration of 4 optical cycles for the same intensity as before.  To reproduce the energy distribution we need to use 1200 Coulomb Sturmian functions per angular momentum. 80 angular momenta are included in the calculations and the non-linear parameter $\kappa$ of the Coulomb Sturmian functions is equal to 0.3.  For this rather stiff problem Arnoldi's algorithm has to include a minimum of 70 Krylov vectors for a time step  $\delta t = 0.05$ a.u.  The calculation takes 24 hours on an 8 processor cluster using  OpenMP.  Fatunla's algorithm also reproduces the same energy distribution but the computer time used is more than four times larger. In fact, we observe that for larger scale problems where the degree of stiffness is important, Fatunla's method requires time steps that become prohibitively small thereby increasing  the computational time.

\subsection{Helium Atom}
\begin{figure}[!ht]
\begin{center}
\includegraphics[width=12cm,height=8cm]{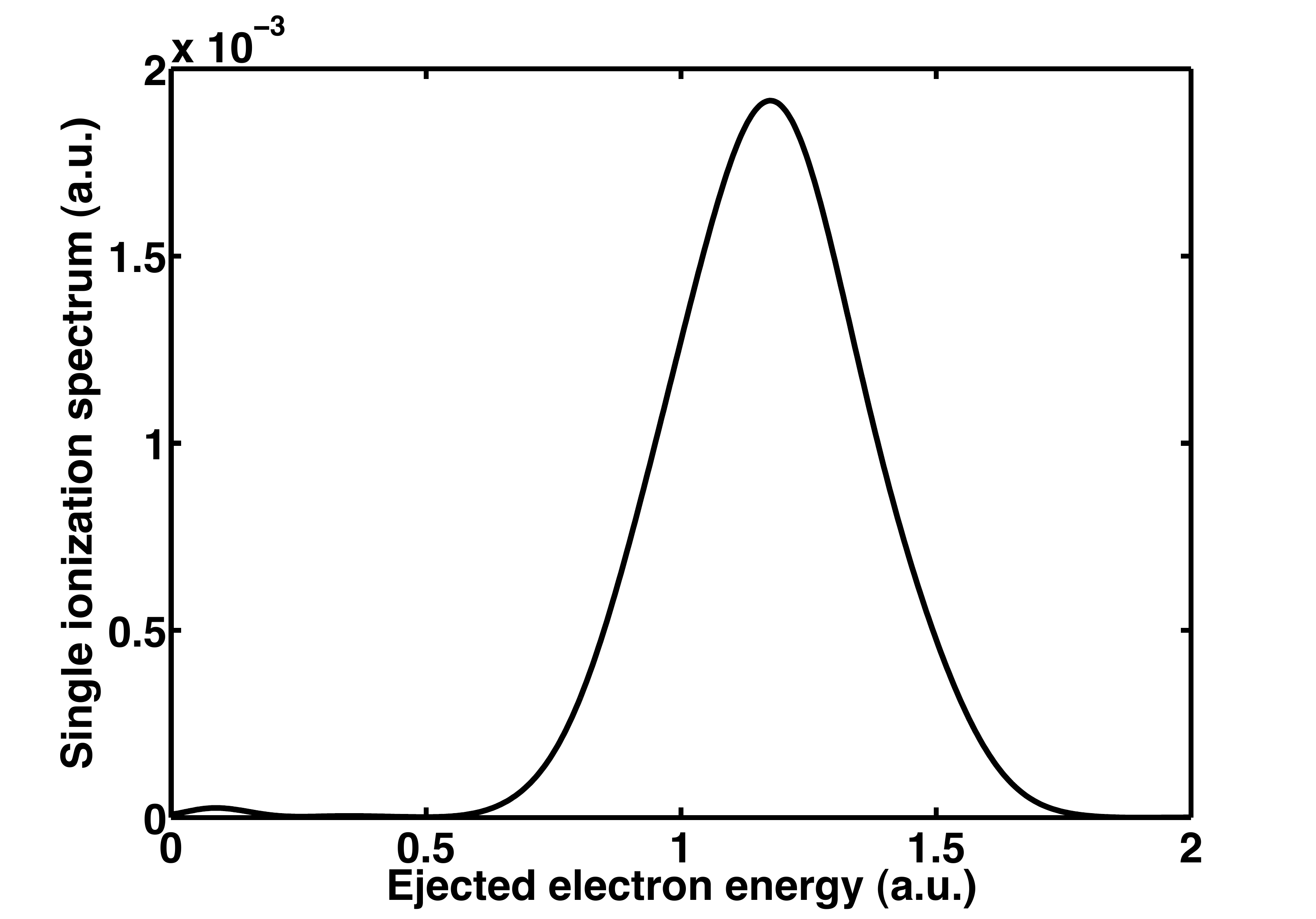}
\end{center}
\caption{Single ionisation spectrum resulting from the interaction of an helium atom with a short cosine square pulse.  Arnoldi's propagator is used. The pulse has a peak intensity of I=10$^{14}$ Watt/cm$^2$, a frequency $\omega=2.1$ a.u. and a duration of 6 optical cycles. The basis-set of functions used is a set of 140 B-spline functions of order 7 per electron angular momentum. The total angular momentum L=0,1,2 and the maximum value of the individual electron angular momentum is three. The box size is 150 a.u.  The Arnoldi propagator uses 40 Krylov vectors.}
\label{fig16}
\end{figure}

In this subsection, we show briefly results for the single ionization of helium by an intense electromagnetic pulse as an example of a more challenging problem. Following the remarks above, we perform here the calculations using only Arnoldi's algorithm. As mentioned before, Fatunla's algorithm is not accurate enough to calculate cross sections in various single ionization channels. The  pulse has a peak intensity I=10$^{14}$ Watt/cm$^2$, a frequency $\omega=2.1$ a.u. and a duration of 6 optical cycles. The wavefunction is expanded in a basis set that uses 140 B-spline functions of order 7 per electron angular momentum  \cite{Bachau}. Three values of the total angular momentum (L=0,1,2) are taken into account and the maximum value of the individual electron angular momentum is three.  The box size is 150 a.u. The step size during the interaction with the pulse is fixed at  0.01 a.u., while after the interaction the propagation used a step size of 1 a.u. The calculations are performed with 40 Krylov vectors.  It takes 31 hours to run on a cluster with 10 Intel Xeon L5520 2.26 GHz processors  using MPI (Message Passing Interface) and 3 GB of RAM per processor. Fig.\ref{fig16} shows the results obtained for the energy distribution of the single ionization of helium.  As expected we observe a dominant peak at 1.2 a.u. which corresponds to the energy conservation. The spectrum is obtained by projecting the wave packet after the end of the pulse on a product of a Coulomb wave of the screened nucleus times a bound state of He$^{+}$.

\subsection{Quantum dot}

In this last section, we consider a different problem where the choice of a very efficient explicit time propagator turns out to be crucial. The system
under consideration is a model for a planar two-electron quantum dot with an anharmonic confining potential. The properties of quantum dots have great resemblance to those of atoms or molecules. Optical lattices, which can be viewed as an array of quantum dots, and well-approved methods from semiconductor physics make quantum dots easily accessible. A confinement of the electrons to a two-dimensional plane is justified, in particular for solid state quantum dots, where the electron gas is localized on a parallel plane between two layers of different semiconductors. The Hamiltonian for this problem is of the form
\begin{equation}
H_{\varepsilon} = {H_1} + {H_2} + {V_{{\mathop{\rm int}} }},\label{eq33}
\end{equation}
where the indices $1$ and $2$ refer to the two electrons.  ${V_{{\mathop{\rm int}} }} =\displaystyle{ \frac{1}{{\,{r_{1\,2}}}}}$, with $ {{r_{1\,2}}}$ being the inter-electronic distance.  The Hamitonians $H_j$ are given by,
\begin{equation}
{H_j} = \frac{1}{2} \mathbf{p}_j^{{\kern 1pt} 2} + \frac{{{\omega ^2}}}{2} \mathbf{r}_j^{{\kern 1pt} 2} + 
\varepsilon {\left( { \mathbf{r}_j^{{\kern 1pt} 2}} \right)^2}, \label{eq34}
\end{equation}
with $\omega$ the harmonic frequency and $\varepsilon$ the strength of the anharmonic perturbation. $\mathbf{r}_j$ and $\mathbf{p}_j$ are the coordinate and momentum of electron $j$, respectively. For $\varepsilon\equiv 0$ our model coincides with the well-known Hooke's atom, which is separable in the centre-of-mass and relative coordinates. The Schr\"odinger equation can be regularized \cite{Schroeter1} using the Jacobian of a suitable parabolic coordinate transformation.  We then write the resulting equation in terms of circular creation and annihilation operators.  A set of selection rules is obtained determining the coupling between basis states and the matrix elements, according to the principal quantum numbers of the harmonic oscillators. The TDSE,
\begin{equation}
H\;\Psi \left( {{{\bf{r}}_{\,1}},{{\bf{r}}_{\,2}},t} \right) = \mathrm{i} \frac{\partial }{{\partial \,t}}\;\Psi \left( {{{\bf{r}}_{\,1}},{{\bf{r}}_{\,2}},t} \right), \label{eq35}
\end{equation}
is solved to obtain $\Psi \left( {{{\bf{r}}_{\,1}},{{\bf{r}}_{\,2}},t} \right)$, with $H$ given in Eq.(\ref{eq33}).
The question of decoherence of these quantum states can be studied through the quantum fidelity, which gives the overlap of the solutions of the TDSE, with and without the potential ${V_{{\mathop{\rm anharmonic}} } =\displaystyle{ \varepsilon {\left(\left( { \mathbf{r}_1^{{\kern 1pt} 2}} \right)^2+\left( { \mathbf{r}_2^{{\kern 1pt} 2}} \right)^2\right)}}}$. The perturbation potential ${V_p =\displaystyle{ \left( { \mathbf{r}_1^{{\kern 1pt} 2}} \right)^2+\left( { \mathbf{r}_2^{{\kern 1pt} 2}} \right)^2}}$ breaks the separability of Hooke's atom. We note that the Hamiltonian $H_{\varepsilon}$ in this case is not explicitly dependent on time and so it is different in nature to the Hamiltonians that we treated in previous examples.
For a general Hamiltonian $H_0$ and a small real parameter $\varepsilon$ that represents the strength of the perturbation, we write
\begin{equation}
{H_\varepsilon } = {H_0} + \varepsilon {\kern 1pt} {V_{{\mathop{\rm p}.} }}\label{eq36}
\end{equation}
The quantum fidelity $F_\varepsilon$ at time $t$ is defined as,
\begin{equation}
{F_\varepsilon }\left( t \right) = {\left| {\left\langle {{\Psi _0}\left( t \right)} \right|\left. 
{{\Psi _\varepsilon }\left( t \right)} \right\rangle } \right|^2},\label{eq37}
\end{equation}
where $\Psi _\varepsilon$ and $\Psi _0$ are the quantum states propagated with Eq.(\ref{eq35}) for a perturbed and non-perturbed Hamiltonian, respectively.  We can expand the quantum fidelity in terms of the perturbation parameter $\varepsilon$ \cite{Gorin}, as,
\begin{equation}
{F_\varepsilon }\left( t \right) = 1 - \chi \left( t \right){\varepsilon ^2} + O\left( {{\varepsilon ^4}} \right),\label{eq38}
\end{equation}
with $ \chi \left( t \right)$ being the quantum susceptibility.  Taking the two first terms, we evaluate ${F_\varepsilon }\left( t \right)$ up to order ${\varepsilon ^2}$, valid near unity.  
\begin{figure}[!b]
\begin{center}
\includegraphics[width=12cm,height=8cm]{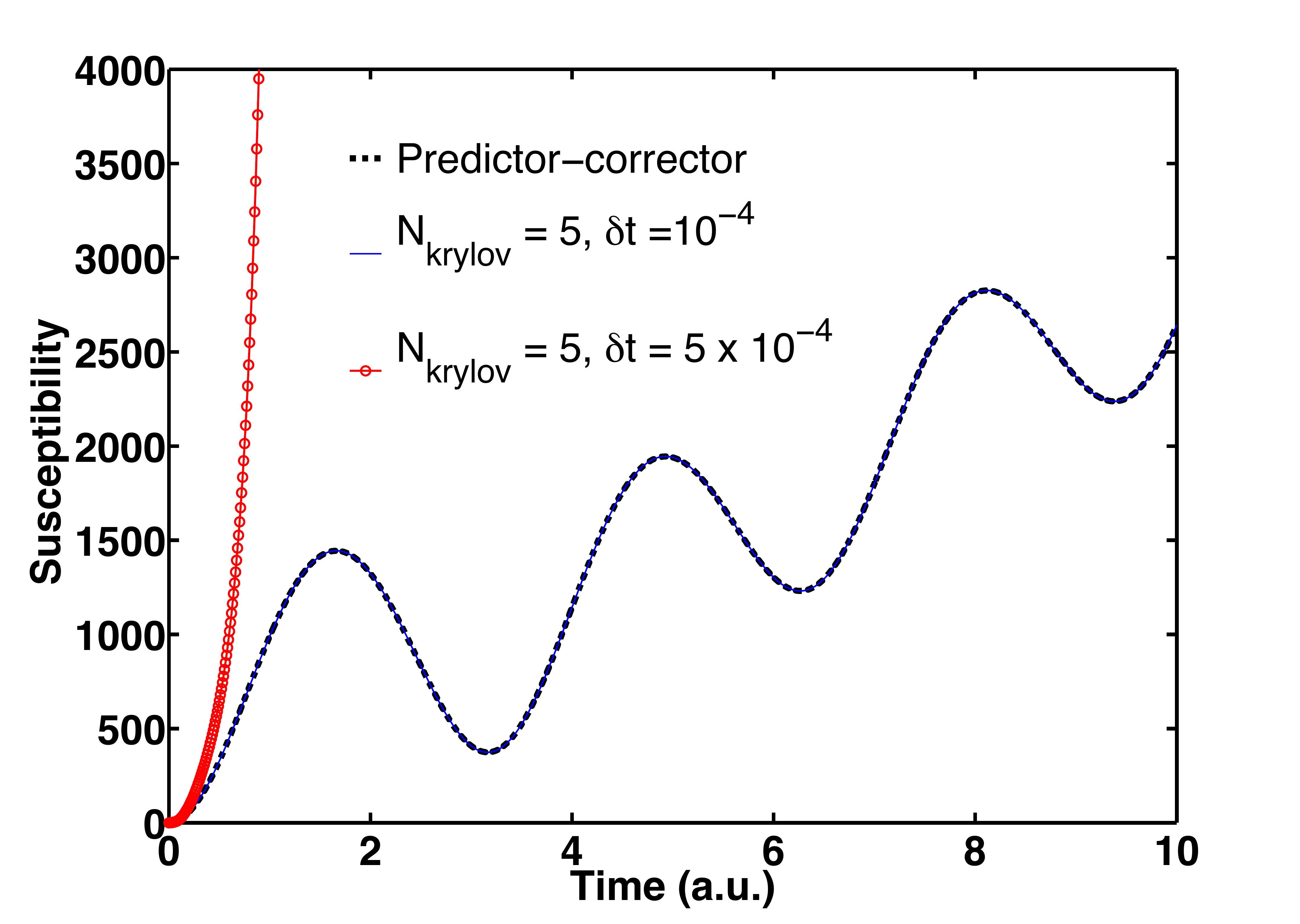}
\end{center}
\caption{(Color online) Susceptibility  $ \chi \left( t \right)$ calculated for a quantum dot with $\omega=$1.0 a.u., using the Arnoldi's propagator. We took 5 Krylov vectors and compared results for two different values of the time step.}
\label{fig17}
\end{figure}
\begin{figure}[!ht]
\begin{center}
\includegraphics[width=12cm,height=8cm]{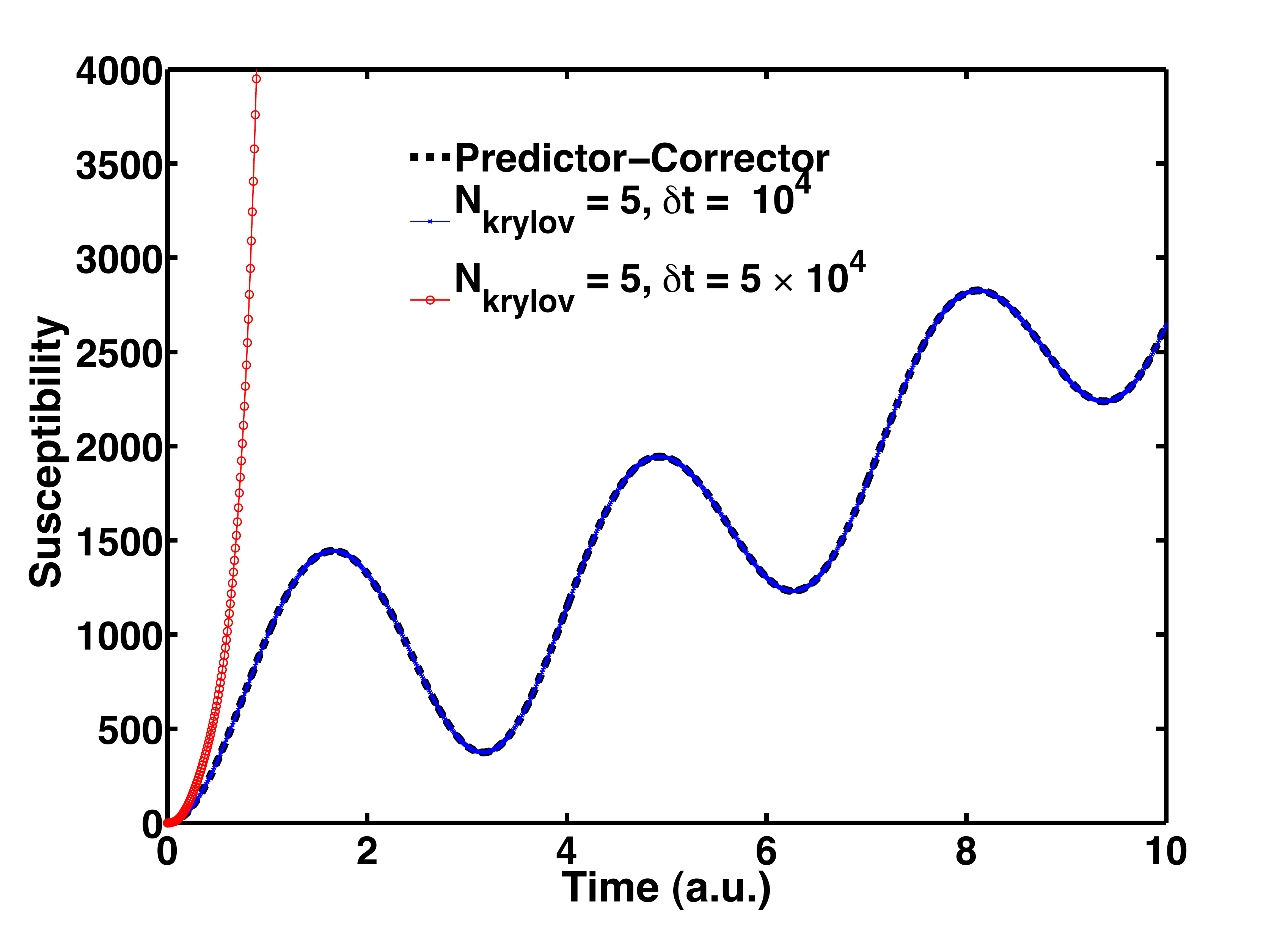}
\end{center}
\caption{(Color online) Susceptibility  $ \chi \left( t \right)$ calculated for a quantum dot with $\omega=$1.0 a.u., using the Arnoldi's propagator. We take two different combinations of time step and of the dimension of the Krylov subspace used, to illustrate the compromise between these two quantities.}
\label{fig18}
\end{figure}
 For our particular case $H_0=H_{\varepsilon=0}$ and consequently $V_{{\mathop{\rm p}} }=\displaystyle{ {\left(\left( { \mathbf{r}_1^{{\kern 1pt} 2}} \right)^2+\left( { \mathbf{r}_2^{{\kern 1pt} 2}} \right)^2\right)}}$.
 The observable calculated in this problem is the susceptibility $\chi \left( t \right)$ and we take the harmonic frequency to be $\omega=$1.0 a.u. and the perturbation parameter to be $\varepsilon=10^{-5}$.  We study the evolution of the initial bound state of energy E=7 a.u. and vanishing angular momentum, singlet state with even parity \cite{Schroeter2}.  The total number of functions in the basis set is 2370.  The integration of the TDSE was first attempted using Fatunla's method.  The stiffness of this problem forces the adaptive time step to become excessively small (of the order of $10^{-5}$) so that the computer time needed by the method becomes of the order of several days instead of seconds. Furthermore the accuracy necessary to represent the effect of very small perturbations on the system could not be achieved. As a consequence we used Arnoldi's integrator, testing different combinations of the values of the time step and of the dimension of the Krylov subspace. It is worth stressing that the time evolution operator calculated within Arnoldi's method is essentially exact since the total Hamiltonian is time independent. However, the stiffness of the problem which is very strong because of the anharmonic character of the potential is expected to impose important constraints on the time step. In Fig.\ref{fig17} we show results for the quantum susceptibility using 5 Krylov vectors and two different time steps.  In order to get converged results, this shows that we need a time step of at least $\delta t = 10^{-4}$ a.u., leading to a computational time of 4 hours. The same calculation performed with the P-C method took 17 days, 8 hours and 29 minutes. Fig.\ref{fig18} shows results for the observable  $\chi \left( t \right)$ under the same conditions as in Fig.\ref{fig17} but using Krylov subspaces of higher dimension ($n_k=7$ and $n_k=9$).  For $n_k=7$ converged results were obtained with a  step size of $\delta t = 5 \times10^{-4}$ a.u. leading to a computational time of 45 minutes.  However this figure illustrates the compromise to be achieved between the size of the time step used and the dimension of the Krylov subspace.  For $n_k=9$, a time step of $\delta t = 10^{-3}$ a.u. leads to a calculation taking 30 minutes of computer time only.  The choice of the optimal value of time step and of the Krylov subspace dimension needs to be balanced.  This means to search for the optimal larger value of the time step for which the propagation will take less iterations.  These calculations performed with the P-C method take  9 days, 17 hours and 14 minutes for $n_k=7$ and $\delta t = 5 \times10^{-4}$ and 8 days, 5 hours and 43 minutes for a.u. $n_k=9$ and $\delta t = 10^{-3}$ a.u. The computer used in these calculations was a single core of a Intel(R) Core (TM) 2 Quad CPU Q 9400(2.66 GHz) with 8 GB main memory.

\section{CONCLUSIONS}

In this contribution, we addressed the problem of the numerical integration of the time-dependent Schr\"odinger equation describing physical processes whose complexity requires the use of state of the art methods. The problem can be reduced to the solution of a system of first order differential equations. The main difficulties we have to face are the size of the system and its stiff character which results from the presence of very high energy eigenvalues in the Hamiltonian spectrum. These difficulties impose important constraints on the choice of the time propagator. Given the size of the system, this time propagator must be explicit. This means that it involves only matrix-vector products instead of solving large system  of algebraic equations at each time step as is the case for implicit methods. In addition, this propagator must have optimum stability and accuracy properties to cope with the stiffness of the system. We have analyzed and compared the performance of two one-step explicit time propagators namely Fatunla's and Arnoldi's algorithms. It turns out that both of these methods share the same optimum stability properties. Nevertheless, we show that their accuracy properties differ significantly in most of the problems that we treat here. As a matter of fact, the accuracy of the method depends essentially on the stiffness of the system to solve which determines the appropriate choice of the propagator. \\

In all the problems considered here, the relative accuracy of Fatunla's method is always  limited to about $10^{-6}$. In some cases, this might be sufficient but we should not forget that when the degree of stiffness increases, the adaptive time step becomes excessively small making the method inapplicable. By contrast, highly accurate results are obtained with Arnoldi's algorithm in all cases treated here. However, for a given time step, there is a minimal number of  Krylov vectors to take into account. If the actual number used is smaller than this minimal number, generally there is an abrupt transition and the results are wrong giving a flat spectrum (in some cases this transition is not so abrupt but is rapid nevertheless.) On the other hand, when the degree of stiffness is high, this minimal number may become very large thereby imposing strong limitations on the applicability of the method. This is the case when the spacing between grid points becomes very small or, for spectral methods, when the size of the basis set is very large. In applying Arnoldi's scheme, it is therefore important to try to reduce the stiffness as much as possible. An obvious way to achieve this is to move to the atomic basis in which the Hamiltonian is diagonal and to eliminate the highest energy eigenvalues which, in principle do not play any physical role. In that case however, the ac-Stark shift of the levels will not be evaluated accurately. In addition, our calculations in the case of the Gaussian potential model clearly show that the energy electron spectrum calculated with Arnoldi's algorithm deteriorates.

\section{ACKNOWLEDGEMENTS}
A.L.F. gratefully acknowledges the financial support of the IISN (Institut Interuniversitaire des Sciences Nucl\'eaires) through contract No. 4.4.504.10, "Atoms, ions and radiation; Experimental and theoretical study of fundamental mechanisms governing laser-atom interactions and of radiative and collisional processes of astrophysical and thermonuclear relevance". F.M.F. and P.F.O'M thank the Universit\'e catholique de Louvain (UCL) for financially supporting several stays at the Institute of Condensed Matter and Nanosciences of the UCL. They also thank The european network COST (Cooperation in Science and Technology) through the Action CM1204 "XUV/X-ray light and fast ions for ultrafast chemistry (XLIC) for  financing one short term scientific mission at UCL. Computational resources have been provided by the supercomputing facilities of the UCL and the Consortium des Equipements de Calcul Intensif en F\'ed\'eration Wallonie Bruxelles (CECI) funded by the Fonds de la Recherche Scientifique de Belgique (F.R.S.-FNRS) under convention 2.5020.11.

\end{document}